\PassOptionsToPackage{dvipsnames}{xcolor}
\documentclass[twocolumn,trackchanges]{aastex631}
\usepackage{newtxtext,newtxmath}
\usepackage[T1]{fontenc}
\usepackage{graphicx}
\usepackage{amsmath}
\usepackage{threeparttable}
\usepackage{booktabs}
\usepackage{multirow}
\usepackage{float}
\usepackage{colortbl}
\usepackage{cleveref}

\newcommand{\RomanNumeralCaps}[1]
    {\MakeUppercase{\romannumeral #1}}

\received{August 5, 2023}
\revised{September 22, 2023}
\accepted{October 3, 2023}
\shorttitle{validating emulated global 21-cm posteriors}
\shortauthors{Dorigo Jones et al.}
\graphicspath{{./}{}}

\begin{document}
\title{Validating posteriors obtained by an emulator when jointly-fitting \\
mock data of the global 21-cm signal and high-z galaxy UV luminosity function}

\author[0000-0002-3292-9784]{J. Dorigo Jones}
\affiliation{Center for Astrophysics and Space Astronomy, Department of Astrophysical and Planetary Sciences, University of Colorado Boulder, CO 80309, USA}
\author[0000-0003-2196-6675]{D. Rapetti}
\affiliation{NASA Ames Research Center, Moffett Field, CA 94035, USA}
\affiliation{Research Institute for Advanced Computer Science, Universities Space Research Association, Washington, DC 20024, USA}
\affiliation{Center for Astrophysics and Space Astronomy, Department of Astrophysical and Planetary Sciences, University of Colorado Boulder, CO 80309, USA}
\author[0000-0002-8802-5581]{J. Mirocha}
\affiliation{Jet Propulsion Laboratory, California Institute of Technology, 4800 Oak Grove Drive, Pasadena, CA 91109, USA}
\affiliation{California Institute of Technology, 1200 E. California Boulevard, Pasadena, CA 91125, USA}
\author[0000-0002-9377-5133]{J. J. Hibbard}
\affiliation{Center for Astrophysics and Space Astronomy, Department of Astrophysical and Planetary Sciences, University of Colorado Boulder, CO 80309, USA}
\author[0000-0002-4468-2117]{J. O. Burns}
\affiliation{Center for Astrophysics and Space Astronomy, Department of Astrophysical and Planetary Sciences, University of Colorado Boulder, CO 80309, USA}
\author[0000-0001-7051-6385]{N. Bassett}
\affiliation{Center for Astrophysics and Space Astronomy, Department of Astrophysical and Planetary Sciences, University of Colorado Boulder, CO 80309, USA}
\correspondingauthor{J. Dorigo Jones}
\email{johnny.dorigojones@colorado.edu}

\begin{abstract}
Although neural-network-based emulators enable efficient parameter estimation in 21-cm cosmology, the accuracy of such constraints is poorly understood. We employ nested sampling to fit mock data of the global 21-cm signal and high-$z$ galaxy ultraviolet luminosity function (UVLF) and compare for the first time the emulated posteriors obtained using the global signal emulator {\tt globalemu} to the `true' posteriors obtained using the full model on which the emulator is trained using {\tt ARES}. Of the eight model parameters we employ, four control the star formation efficiency (SFE), and thus can be constrained by UVLF data, while the remaining four control UV and X-ray photon production, and the minimum virial temperature of star-forming halos ($T_{\rm min}$), and thus are uniquely probed by reionization and 21-cm measurements. For noise levels of 50 and 250 mK in the 21-cm data being jointly-fit, the emulated and `true' posteriors are consistent to within $1\sigma$. However, at lower noise levels of 10 and 25 mK, {\tt globalemu} overpredicts $T_{\rm min}$ and underpredicts $\gamma_{\rm lo}$, an SFE parameter, by $\approx3-4\sigma$, while the `true' {\tt ARES} posteriors capture their fiducial values within $1\sigma$. We find that jointly-fitting the mock UVLF and 21-cm data significantly improves constraints on the SFE parameters by breaking degeneracies in the {\tt ARES} parameter space. Our results demonstrate the astrophysical constraints that can be expected for global 21-cm experiments for a range of noise levels from pessimistic to optimistic, and also the potential for probing redshift evolution of SFE parameters by including UVLF data.
\end{abstract}

\section{Introduction}
\label{sec:intro}
A promising tool for probing the physics of the early Universe is the 21-cm cosmological signal arising from the neutral hydrogen gas that permeated the intergalactic medium (IGM) before, during, and after the formation of the first stars and galaxies (\citealt{Madau97}; for reviews see \citealt{Furlanetto06, Bera23}). The spin-flip transition in neutral hydrogen emits low-frequency radiation at 1420.4 MHz ($\lambda \approx 21$ cm), which has been redshifted to low radio frequencies ($\nu \lesssim 200$ MHz, corresponding to redshifts $z \gtrsim 6$) due to cosmic expansion and encodes the high-redshift evolution of the IGM. The 21-cm signal has both an anisotropic component (power spectrum) and an isotropic, sky-averaged component (global signal; \citealt{Shaver99}), whose brightness temperature is measured as a differential temperature relative to the Cosmic Microwave Background (CMB) radiation.

An unambiguous detection of the global 21-cm signal has the potential to reveal the true astrophysical and cosmological properties associated with the Dark Ages ($z > 30-40$), Cosmic Dawn (CD; $10 \lesssim z \lesssim 40$), and the Epoch of Reionization (EoR; ending by $z \approx 6$). However, the global 21-cm signal is particularly difficult to detect due to the presence of significant foreground emission from the Milky Way that is $4-6$ orders of magnitude brighter than the underlying signal, making a robust Bayesian forward modelling approach necessary to properly recover and exploit the global 21-cm signal (e.g., \citealt{Bernardi16, Liu20, Shen22}).

Radio telescopes on Earth have provided some constraints on the 21-cm power spectrum (e.g., \citealt{GMRT, LOFAR, MWA, LEDA, HERA}) and global 21-cm signal (e.g., \citealt{EDGES, SARAS2, SARAS3}). The claimed EDGES detection has been met with skepticism (see e.g., \citealt{Hills18, Bradley19, Tauscher20, SimsPober20}) particularly because of the systematics involved with measuring the global signal and recently because it has been found to be in tension with the non-detection published by SARAS 3 \citep{SARAS3}. To properly recover the underlying global 21-cm signal, the beam-weighted foreground (i.e., foreground emission convolved with the antenna beam) and instrumental systematics must be carefully fitted and removed (e.g., \citealt{PaperII, Hibbard20, PaperIV, Pagano22, Murray22, Anstey23, Hibbard23}). Radio frequency interference (RFI) is a large systematic due to artificial and ionospheric terrestrial contamination which can be avoided by measuring the 21-cm signal from the pristine radio environment of the far side of the Moon. Upcoming NASA Commercial Lunar Payload Services (CLPS) missions ROLSES (2023, at the lunar south pole; \citealt{Burns21}) and LuSEE-Night (early 2026, on the far side; \citealt{LuSEENight}) will lay the path for future lunar far side radio telescope arrays capable of measuring the 21-cm global signal and power spectrum (e.g., FARSIDE \citep{FARSIDEBurns21} and FarView \citep{FarViewPoster}).

Physically-motivated models for the global 21-cm signal have various astrophysical and cosmological parameters that affect the shape of the signal. Multiple studies have attempted to constrain such model parameters when fitting a measured global 21-cm signal via a Bayesian, likelihood-based approach (e.g., \citealt{Monsalve18, Mirocha19, Monsalve19, Qin20, Bevins22a, Bevins23}). In this work, we perform a similar Bayesian parameter estimation analysis for eight astrophysical parameters using the publicly available model {\tt ARES} (Accelerated Reionization Era Simulations\footnote{\url{https://github.com/mirochaj/ares}; v0.9; git commit hash: fd77c4a86982d25fdad790d717f8bf5eecff4eb8}; \citealt{Mirocha14, Mirocha17}) by fitting mock data of the global 21-cm signal and numerically sampling the full posterior distribution of these parameters via nested sampling. We examine the improvement in constraining power on these parameters when jointly-fitting mock data of the high-$z$ galaxy rest-frame ultraviolet (UV) luminosity function (LF) in addition to the global 21-cm signal. We present the first nested sampling constraints on {\tt ARES} parameters when fitting a mock global 21-cm signal and UVLF that are calibrated to real UVLF data. In doing so, we forecast the level of astrophysical constraints that can be expected for different noise levels of global 21-cm experiments in combination with UVLF data.

The recent development of neural-network-based emulators for the global 21-cm signal, such as {\tt globalemu} (\citealt{globalemu}, v1.8.0, Zenodo, \href{https://doi.org/10.5281/zenodo.8178850}{doi:10.5281/zenodo.8178850}), {\sc 21cmVAE} \citep{21cmVAE}, and {\sc 21cmEMU} \citep[][which also emulates other quantities such as the 21-cm power spectrum and the UVLF]{Breitman23}, enables fast, efficient parameter estimations when fitting the global signal. To our knowledge, there is currently no study that shows a direct comparison of the parameter estimates obtained when using an emulator versus the corresponding full model of the global 21-cm signal in the likelihood. The accuracy of an emulator is determined by computing the root mean squared error (RMSE) between model (i.e. simulated) and network (i.e. predicted) data realizations in a test set, while a fully Bayesian parameter inference and model comparison analysis is much more computationally demanding and yields a formal comparison of the posteriors \citep{Trotta08}.

Parameter estimation using a full model of the global signal in the likelihood is computationally expensive for most existing models. Most global 21-cm signal models are semi-numerical and generate a realization of the signal on the order of minutes to hours \citep{Thomas09, Santos10, Mesinger11, FialkovBarkana14, Ghara15, Ghara18, Murray20, Schneider23, Schaeffer23, Hutter23}, which hinders the ability to perform an analysis that requires on the order of 10$^5$ likelihood evaluations. In contrast, the semi-analytical code {\tt ARES} generates a realization of the global 21-cm signal on the order of seconds, owing its speed primarily to the fact that it evolves the mean radiation background directly as opposed to averaging over large cosmological volumes. Therefore, we use {\tt ARES} in a Bayesian nested sampling analysis to obtain the `true' posterior distributions and for the first time directly compare them to the emulated posteriors from {\tt globalemu}.

We generate the mock global 21-cm signal and high-$z$ UVLF using {\tt ARES} with fiducial parameter values that are calibrated to the \citet{Bouwens15} UVLF at $z=5.9$ \citep{Mirocha17}. We emphasize that the basic {\tt ARES} UVLF model we employ accurately fits UVLFs at $z\approx6-10$ obtained by either {\it HST} or {\it JWST} \citep{Mirocha23}, and so our results would not change if we were to fit mock data calibrated to newer {\it JWST} UVLF measurements at these redshifts. However, given early indications of a departure from the predictions of {\it HST}-based models at $z\gtrsim10$ (see, e.g., \citealt{Naidu22, Lovell23, Donnan23, Finkelstein23, Harikane23, Mason23, Boylan-Kolchin23, Bouwens23}), fitting {\it JWST} UVLFs at $z\gtrsim10$ would require non-trivial changes to the UVLF model we employ \citep{Mirocha23}. We defer such analysis to future work (see also \citealt{Zhang22}).

To summarize, we pursue three main goals: (1) numerically sample the full posterior distribution of eight astrophysical parameters in {\tt ARES}, which control the star formation efficiency and UV and X-ray photon production per unit star formation in galaxies, when fitting mock global 21-cm signal data with varying noise levels; (2) validate and examine the accuracy of the posteriors obtained by our version of the publicly available neural network emulator {\tt globalemu} that we trained with {\tt ARES}; and (3) study the constraints from jointly-fitting high-$z$ galaxy UVLF mock data along with the simulated global 21-cm signal.

In Section~\ref{sec:analysis}, we describe our methods for obtaining marginalized posterior distributions via nested sampling when fitting mock data of the global 21-cm signal and UVLF. We also describe the training of the {\tt globalemu} neural network and the generation of the mock data being fit. In Section~\ref{sec:results}, we present the results from nested sampling analyses, primarily comparing the posteriors obtained when using the emulator {\tt globalemu} in the likelihood versus the full model {\tt ARES}, and also examining the effect on posteriors when jointly-fitting with the high-$z$ galaxy UVLF mock data. Finally, we summarize our results and conclusions in Section~\ref{sec:conclusions}.

\section{Analysis}
\label{sec:analysis}
In this section, we describe our analysis method for obtaining the posterior distributions for eight astrophysical parameters in {\tt ARES} when fitting a mock global 21-cm signal plus statistical noise. The main steps to define our Bayesian analysis are: (1) selecting a sampling method, (2) selecting a fiducial model for the global 21-cm signal, and (3) generating mock data by adding to the simulated global signal a noise realization at a statistical error level corresponding to a given integration time. We also train a neural network to emulate the {\tt ARES} global signal model and study its accuracy versus the full {\tt ARES} model in producing realizations of the signal.

Note that for this work, we are not concerned with systematic uncertainties such as the beam-weighted foreground, radio frequency interference (RFI; either from terrestrial contamination or the instrument), and environmental horizon and surface conditions (for studies on such effects, see e.g., \citealt{SARAS2, Kern20, Bassett20, Hibbard20, Bassett21, Pagano22, Leeney22, Murray22, Anstey23, Hibbard23}).

\subsection{Likelihood}
\label{subsec:likelihood}
Bayesian inference allows us to estimate the posterior distribution $P(\theta | \boldsymbol{D}, m)$ of a set of parameters $\theta$ in a model $m$, given observed data $\boldsymbol{D}$ with priors $\pi$ on the parameters (also written $P(\theta | m)$). This is achieved via the Bayes' theorem:
\begin{equation}
\label{eqn:Bayes}
P(\theta | \boldsymbol{D}, m) = \frac{\mathcal{L} (\theta) \pi (\theta)}{{\it Z}},
\end{equation}

\noindent where $\mathcal{L}$ is the likelihood function, or the probability of the data given the parameters of the model (also written $P(\boldsymbol{D} | \theta, m)$), and the normalizing factor {\it Z} is the Bayesian evidence, or marginal likelihood over the priors (also written $P(\boldsymbol{D} | m)$), which can be used for model comparison.

For all of the fits performed in this paper, we sample from a multi-variate log-likelihood function assuming Gaussian-distributed noise:
\begin{equation}
\label{eqn:likelihood}
\log \mathcal{L}(\theta) \propto [\boldsymbol{D} - m(\theta)]^T \boldsymbol{C}^{-1} [\boldsymbol{D} - m(\theta)],
\end{equation}

\noindent where $\boldsymbol{C}$ is the noise covariance matrix of the data, which we assume to be diagonal. In this paper, we fit mock data realizations for the global 21-cm signal ($\boldsymbol{D_{21}}$) and the UVLF ($\boldsymbol{D_{\rm UVLF}}$) instead of real data, although for the latter, the mock data are calibrated to real measurements of the high-$z$ galaxy UVLF (see Section~\ref{subsec:mockdata}). Hence, we know the input, or fiducial, values of the parameters whose posteriors we numerically sample and can evaluate the validity of the sampling methods and the accuracy of the {\tt ARES} model and {\tt globalemu} emulator based on the expectation of marginalized posterior distributions around the fiducial parameter values.

\subsection{Combined Constraints}
\label{subsec:combinedconstraints}
To better realize the constraints that are achievable from global 21-cm signal experiments, in addition to fitting only the mock global signal, we also perform joint-fits that combine the model constraining powers from the global signal and high-$z$ galaxy UVLF mock data. Using Equation~\ref{eqn:likelihood}, we construct separate log-likelihood functions for the global 21-cm signal and the UVLF. For the joint-fits, we form a log-likelihood by adding both individual likelihoods (see, e.g., \citealt{Chatterjee21, Bevins23}):
\begin{equation}
\label{eqn:jointL}
\log \mathcal{L}_{\rm joint} = \log \mathcal{L}(\boldsymbol{D_{21}}|\theta) + \log \mathcal{L}(\boldsymbol{D_{\rm UVLF}}|\theta).
\end{equation}

We evaluate the separate log-likelihood functions at the same set of parameters using the same priors to sample the full posterior distribution, as the models we employ for the 21-cm signal, $m_{21}(\theta)$, and for the UVLF, $m_{\rm UVLF}(\theta)$, are both generated using the {\tt ARES} framework (see Section~\ref{subsec:model}).\footnote{In Equation~\ref{eqn:jointL}, we set the relative weights for the log-likelihoods equal to unity. It is also possible to explore weights by which to multiply the separate log-likelihoods when combining data sets from different experiments (e.g., \citealt{Lahav00, Hobson02}); we leave consideration of such analysis for future work.} For the global 21-cm signal likelihood, the noise covariance matrix $\boldsymbol{C_{21}}$ is a diagonal array of constant values corresponding to the square of the estimated noise level $\sigma_{21}$. For the UVLF likelihood, the main diagonal elements of $\boldsymbol{C_{\rm UVLF}}$ are the same as the errors on the $z=5.9$ UVLF data by \citet{Bouwens15} (see Section~\ref{subsec:mockdata}).

\subsection{Nested Sampling}
\label{sec:nestedS}

We employ the Bayesian inference method of nested sampling (\citealt{Skilling04}; for reviews see \citealt{Ashton22, Buchner23}). Conceptually, nested sampling algorithms converge on the best parameter estimates by iteratively removing regions of the prior volume with lower likelihood. Nested sampling computes both the evidence and posterior samples simultaneously (by recasting the multi–dimensional evidence integral into a one–dimensional integral), whereas Markov Chain Monte Carlo (MCMC) samplers calculate only the posterior.

In general, Monte Carlo methods like nested sampling and MCMC are computationally expensive because they require many likelihood evaluations to sample the converged posterior distributions. We choose nested sampling instead of MCMC because the former is designed to better constrain complex parameter spaces with “banana”-shaped curved degeneracies and/or multi-modal distributions \citep{Buchner23}. Another likelihood-based method that has been applied to parameter estimation of the global 21-cm signal is Fisher-matrix analysis \citep{Liu13, Munoz20, Hibbard22, Mason23Fish}, which assumes multi-variate Gaussian posterior distributions and requires only $\mathcal{O}(N)$ likelihood evaluations for $N$ parameters being sampled. Fisher analysis is efficient but provides an accurate description only when the posteriors are symmetric, Gaussian, and uni-modal (e.g., \citealt{Trotta08, Ryan23}); however, Fisher matrix generalizations exist \citep{Heavens16} such as adding higher order matrices \citep{DALI}. There are also `likelihood-free' inference methods (also called simulation-based inference, see \citealt{Cranmer20}), which have been shown to provide accurate posteriors at a relatively low computational cost \citep{PrelogovicMesinger23}.

Two nested sampling algorithms in particular have been used in 21-cm cosmology and have been shown to efficiently sample posterior distributions: {\tt MultiNest} (\citealt{Feroz08, Feroz09, Feroz19}) and {\tt PolyChord} (\citealt{Handley15a, Handley15b}, v1, Zenodo, \href{https://doi.org/10.5281/zenodo.3598030}{doi:10.5281/zenodo.3598030}). In both {\tt MultiNest} and {\tt PolyChord}, an initial number of `live' points, $n_{\rm live}$, are generated in the prior volume, which are used to eventually converge on the best parameter estimates, but the two nested sampling algorithms differ in how they replace live points. For a more in-depth comparison of {\tt MultiNest} and {\tt PolyChord} see Section 2 of \citet{Lemos23}. Because of their different approaches for replacing live points, {\tt MultiNest} and {\tt PolyChord} are known to perform differently depending on the number of dimensions, or the number of parameters being constrained (see Fig. 4 in \citealt{Handley15a}). We primarily utilize {\tt MultiNest} for our analyses, and we show for one joint-fit that the two nested sampling algorithms converge on roughly the same result for the same $n_{\rm live}$ but with {\tt MultiNest} being much more efficient than {\tt PolyChord} for constraining eight astrophysical parameters in {\tt ARES} (see Section~\ref{subsec:joint}).

\subsection{Modeling the Global 21-cm Signal and UVLF}
\label{subsec:model}

\begin{table*}
    \caption{Astrophysical parameters in {\tt ARES} to be fit with mock global 21-cm signal and high-$z$ UVLF data}
    \begin{center}
    \begin{tabular}{cccc}
    \toprule
    \addlinespace
        Parameter & Description & Prior range (with units) & Fiducial value, $\theta_0$\\
    \addlinespace
    \hline
    $c_X$ & normalization of X-ray luminosity -- SFR relation & Log unif. [$10^{36}$, $10^{41}$]erg s$^{-1}$ (M$_{\odot}$ yr$^{-1}$)$^{-1}$ & $2.6 \times 10^{39}$ \\ %$10^{39.415}$ \\
    $f_{\rm esc}$ & escape fraction of UV photons & Uniform [0,1] & 0.2 \\
    $T_{\rm min}$ & minimum virial temperature of star-forming halos & Log unif. [3 $\times 10^2$, 5 $\times 10^5$]K & $10^4$\\
    $\log N_{\rm H \RomanNumeralCaps{1}}$ & neutral hydrogen column density in galaxies & Uniform [18,23] & 21 \\
    $f_{\rm \star,0}$ & peak star formation efficiency & Log unif. [10$^{-5}$, 10$^0$] & $0.05$ \\ %10^{-1.301}
    $M_{\rm p}$ & dark matter halo mass at $f_{\rm \star,0}$ & Log unif. [10$^8$, 10$^{15}$]$M_{\odot}$ & $2.8 \times 10^{11}$ \\%$10^{11.447158}$\\
    $\gamma_{\rm lo}$ & low-mass slope of $f_{\rm \star} (M_{\rm h})$& Uniform [0,2] & 0.49\\
    $\gamma_{\rm hi}$ & high-mass slope of $f_{\rm \star} (M_{\rm h})$ & Uniform [-4,0] & -0.61\\
    \bottomrule
    \end{tabular}
    \end{center}
    \label{tab:params}
\end{table*}

\begin{figure}
    \includegraphics[scale=0.42]{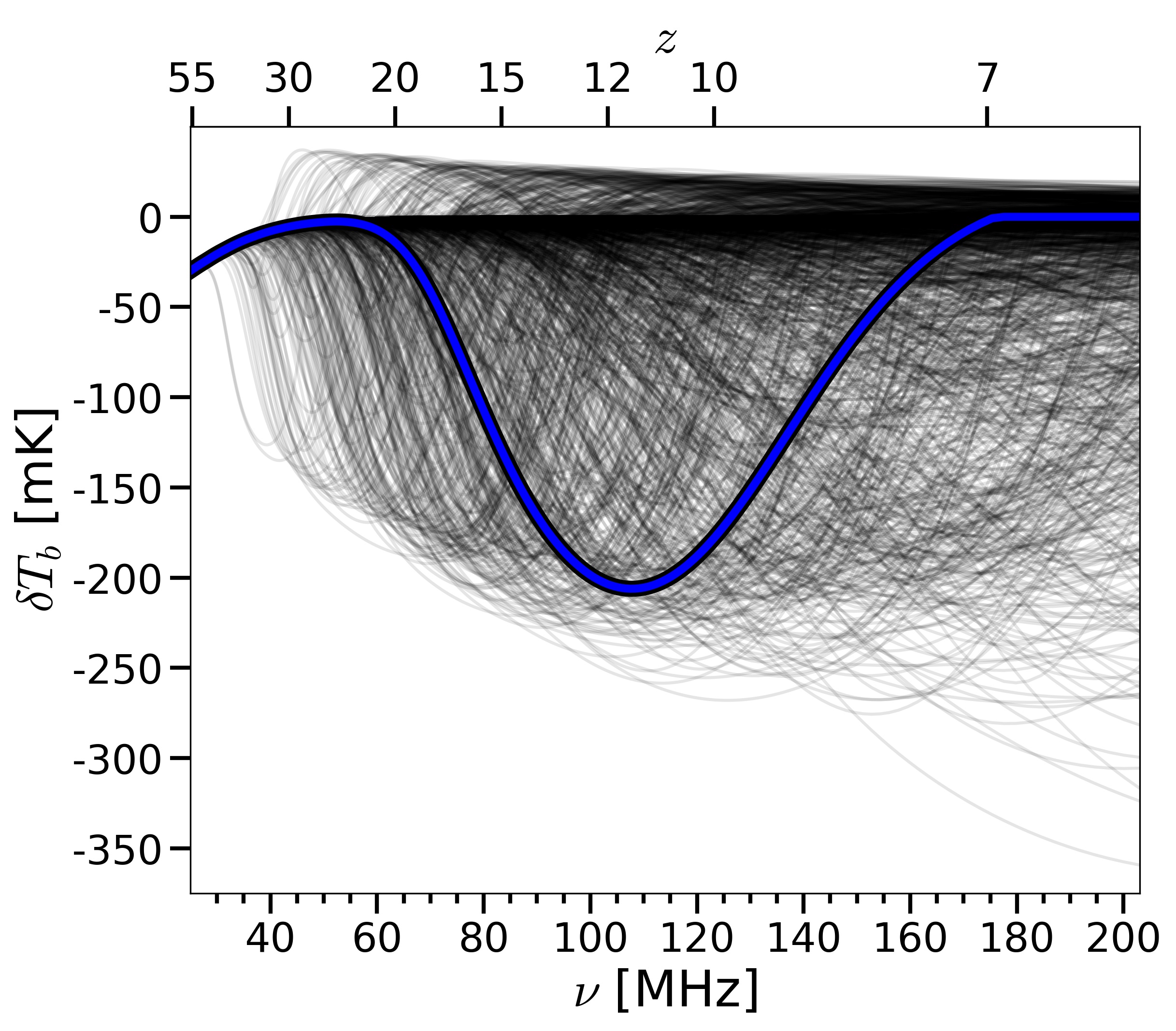}
    \caption{Representative subset of the training set (10\% out of 24,000 total) containing mock global 21-cm signals generated by {\tt ARES} when varying eight astrophysical parameters. The full training set was used to train {\tt globalemu} (see Table~\ref{tab:params} for the parameter ranges). Shown in bolded blue is the fiducial global 21-cm signal to which we add Gaussian-distributed noise at different levels to form the mock 21-cm data sets that we fit. The mock UVLF data that we add in our joint-fits are also generated by {\tt ARES} using the same fiducial parameter values (see Table~\ref{tab:params}) that were obtained via calibration to the \citet{Bouwens15} $z=5.9$ UVLF by \citet{Mirocha17}, as described in Section~\ref{subsec:mockdata}.}
    \label{fig:trainingset}
\end{figure}

\begin{figure}
    \includegraphics[scale=0.42]{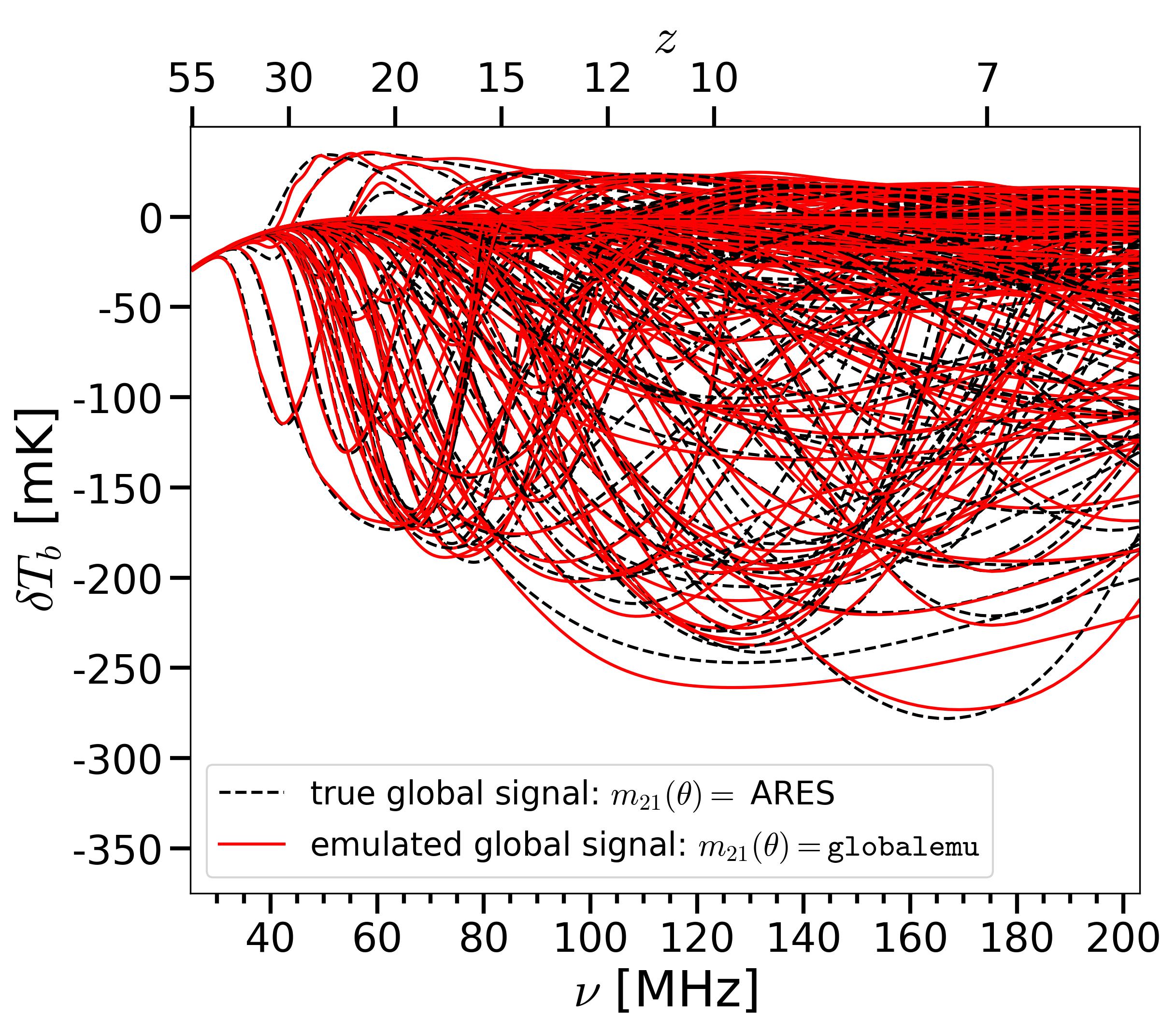}
    \includegraphics[scale=0.42]{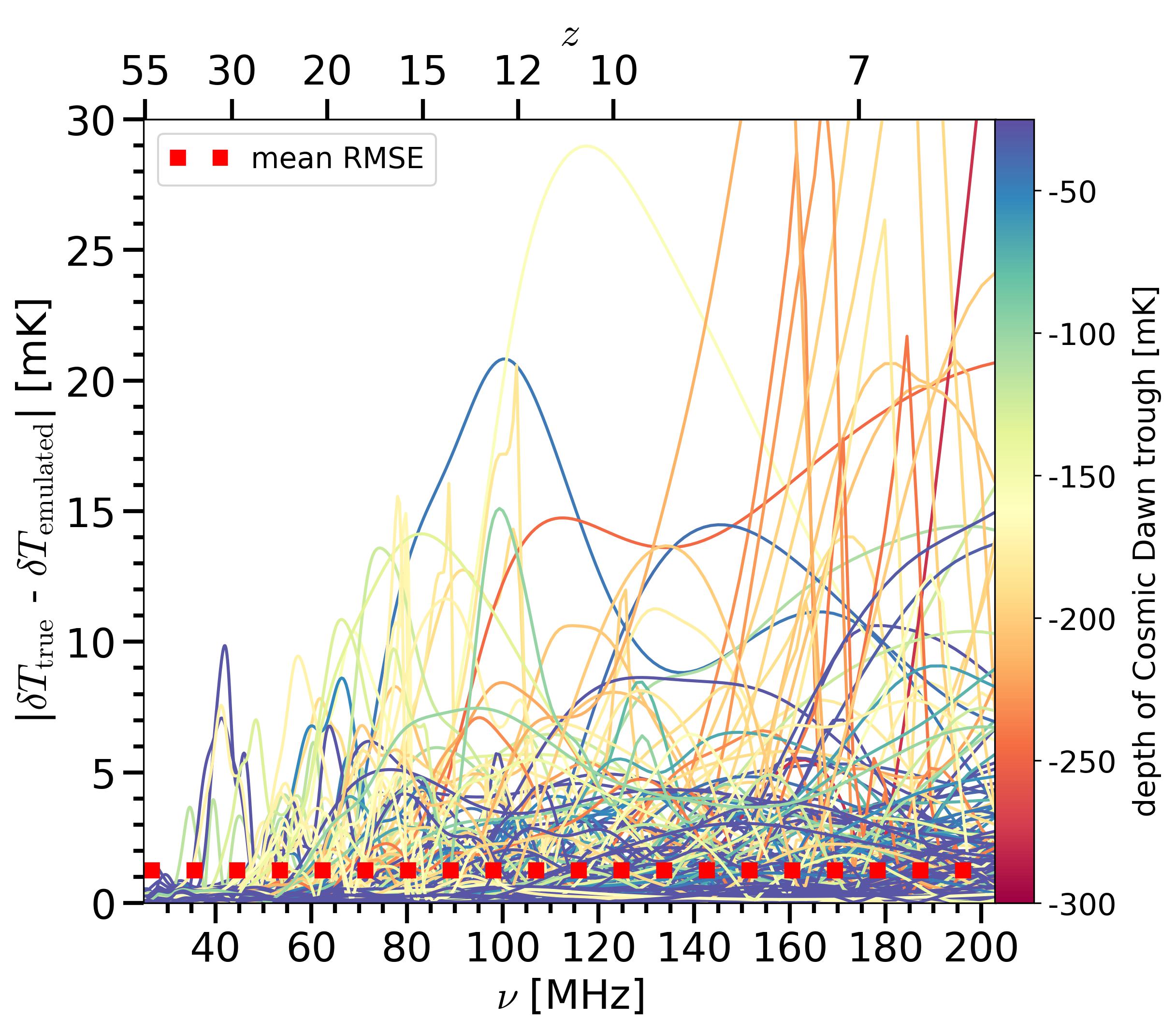}
    \caption{{\it Top:} Representative subset of the test set (200 out of 2,000) generated by {\tt ARES} (`true' global signals; black, dashed curves) and the corresponding subset of emulations from the {\tt globalemu} network (solid, red curves) trained on the {\tt ARES} training set using the architecture [32, 32, 32]. {\it Bottom:} Differences between the emulated and `true' signals in the top panel (i.e., emulation residuals), with color depicting the depth of the Cosmic Dawn (CD) trough of the respective signal. The horizontal dotted, red line indicates the mean RMSE of 1.25 mK between the emulated and `true' signals in the full test set (see Section~\ref{subsec:emulator}).}
    \label{fig:testset}
\end{figure}

To simulate the global 21-cm signal (and high-$z$ galaxy UVLF), we use the physically-motivated, semi-analytical code {\tt ARES}, which is the union of a 1D radiative transfer code developed in \citet{Mirocha12} and a uniform radiation background code described in \citet{Mirocha14}. {\tt ARES} outputs realizations of the global 21-cm signal and galaxy LF in just seconds, which makes it computationally feasible to perform direct parameter estimation using the full model {\tt ARES} rather than an emulator in the likelihood of a nested sampling analysis.\footnote{Each fit using the full {\tt ARES} model was performed on the Blanca compute cluster operated by University of Colorado Research Computing, employing three nodes with 30 CPU cores per node (parallelized via Intel/IMPI compiler) and utilizing $\sim$100 GB of total memory.} Although {\tt ARES} contains cosmological parameters that affect the shape of the Dark Ages trough, in this work we focus on demonstrating the astrophysical constraints that are achievable when fitting the Cosmic Dawn and reionization redshift ranges.

For high-$z$ galaxies, the observed LF probes the rest-frame UVLF, $\phi(M_{\rm UV})$, and so the UVLF model primarily depends on the star formation rate of massive, young stars. The {\tt ARES} model is motivated by studies of the high-$z$ galaxy LF based on abundance matching, and the fiducial model ignores dust extinction (which has a minor impact on the conversion between the observed and intrinsic LF at $z \gtrsim 6$) and suggested redshift evolution of the star formation efficiency (SFE). {\tt ARES} assumes a multi-color disk (MCD) spectrum for high-mass X-ray binaries (HMXBs; \citealt{Mitsuda84}) and uses the BPASS version 1.0 single-star models for continuous star formation to derive the UV photon production efficiency \citep{Eldridge09}.

For full descriptions of how {\tt ARES} models the galaxy UVLF and the global 21-cm signal, see Section 2 of \citet{Mirocha17}. Here we will provide a brief description of the UVLF model to highlight the SFE parametrization. The two components required to calculate the UVLF are (1) the intrinsic luminosity $L$ of galaxies as a function of dark matter (DM) halo mass $M_{\rm h}$, and (2) the DM halo mass function (HMF) (i.e., the number of DM halos per mass bin per co-moving volume of the Universe). The HMF has been well-studied (e.g., \citealt{PressSchechter, Bond91, Murray13}), and in {\tt ARES} it is calculated {\it a priori} in lookup tables using an analytical construct that assumes halos form by spherical collapse. The luminosity of each halo can be written in terms of the star formation rate, which is itself the product of the SFE, $f_{\rm \star}$, and the baryon mass accretion rate (MAR). The MAR is derived directly from the HMF (see, e.g., \citealt{Furlanetto17, Mirocha21}), and so all that is needed to calculate the UVLF is a parametrization for the SFE. Here, as in \citet{Mirocha17}, we assume the SFE is a double power law in $M_{\rm h}$:
\begin{equation}
\label{eqn:SFE}
f_{\rm \star} (M_{\rm h}) = \frac{f_{\rm \star,0}}{\left(\frac{M_{\rm h}}{M_{\rm p}} \right)^{\gamma_{\rm lo}} + \left(\frac{M_{\rm h}}{M_{\rm p}} \right)^{\gamma_{\rm hi}}},  
\end{equation}

\noindent where $f_{\rm \star,0}$ is the peak SFE at mass $M_{\rm p}$, and $\gamma_{\rm lo}$ and $\gamma_{\rm hi}$ are the power-law indices at low and high masses, respectively.

We sample the full posterior distribution of eight parameters, including the four SFE parameters ($f_{\rm \star,0}$, $M_{\rm p}$, $\gamma_{\rm lo}$, and $\gamma_{\rm hi}$) and four other astrophysical parameters: $c_X$, $f_{\rm esc}$, $T_{\rm min}$, and $\log N_{\rm H \RomanNumeralCaps{1}}$. The production and release of X-ray photons in galaxies is controlled by $c_X$ and $\log N_{\rm H \RomanNumeralCaps{1}}$; the escape of UV photons is controlled by $f_{\rm esc}$; and the minimum virial temperature which determines the number of collapsed star-forming halos is controlled by $T_{\rm min}$. In Table~\ref{tab:params}, we summarize these eight parameters and give the flat prior ranges used in the nested sampling analyses and also when training the {\tt globalemu} network on {\tt ARES} mock global 21-cm signals. The flexible {\tt ARES} parameter space allows us to set wide, uninformative priors over these free parameters that are still physically meaningful. In order to facilitate a complete exploration of the prior volume, for four parameters, $c_X$, $T_{\rm min}$, $f_{\rm \star,0}$, and $M_{\rm p}$, we sample from their prior ranges uniformly in log10-space, as shown in Table~\ref{tab:params}. Our prior ranges are centered on some empirically-motivated values (see Section~\ref{subsec:mockdata} for description of fiducial parameter values), but we give multiple orders of magnitude on either side of those values to accommodate potentially dramatic departures at high-$z$ and to capture the full resulting converged posterior distributions (see Section~\ref{sec:results}).

One of our main goals is to directly compare the posteriors when using an emulator for the global 21-cm signal versus when using the full model on which the emulator was trained. In the next sub-section, we describe the construction of the training set for the emulator and directly assess the accuracy of the emulated signals compared to the `true,' input ones.

\vspace{10mm}
\subsection{Emulating {\tt ARES} with {\tt globalemu}} 
\label{subsec:emulator}

We employ the publicly available global 21-cm signal emulator {\tt globalemu} \citep{globalemu} for our analyses, though other emulators for the signal do exist such as {\sc 21cmVAE} \citep{21cmVAE}, {\sc 21cmGEM} \citep{Cohen20}, and the recently released {\sc 21cmEMU} \citep{Breitman23}; we leave a comparison of the posteriors obtained from different global 21-cm signal emulators to future work. To obtain a trained {\tt globalemu} neural network that accurately emulates {\tt ARES}, we first create a large training set of simulated global 21-cm signals generated by {\tt ARES} and then train {\tt globalemu} on this training set. For the latter step, we test multiple network architectures (i.e., different numbers of nodes and hidden layers composing the network; see \citealt{globalemu} for a detailed description of the network).

%\vspace{-17mm}
To create the training set, we generate global 21-cm signals from {\tt ARES} by drawing random values\footnote{At this time, we do not impose constraints on the CMB optical depth, $\tau_e$, or the neutral hydrogen fraction, $x_{\rm H I}$, when creating the training sets. This results in a number of unphysical signals with $\tau_e > 1\sigma$ from the value obtained by \citet{Planck18} and/or $x_{\rm H I} \gtrsim 5\%$ at $z=5.3$, despite the signals being generated from physically reasonable parameter ranges. The existence of unphysical signals in our training set is thus a contaminating factor toward constraining actual data, which is beyond our scope of testing the accuracy of the {\tt globalemu} emulation of {\tt ARES}.} from the parameter ranges given in Table~\ref{tab:params}. Each signal spans the redshift range $z = 6-55$ with a redshift spacing of $\delta z=0.1$, similar to \cite{Bevins22a}. The training set that is ultimately used to train the {\tt globalemu} network used for analyses presented in this work contains 24,000 mock signals. A representative subset of this training set is shown in Figure~\ref{fig:trainingset}. We also generated training sets of sizes 5,000, 10,000, and 20,000, which all resulted in less accurate trained networks. The marginal improvement of 10\% in the RMSE of the resulting trained network obtained when using a training set of size 24,000 compared to 20,000, however, indicates that increasing the number of global signals in the training set above 24,000 would not significantly affect our results. In addition, we also created a so-called test set of 2,000 global signals using {\tt ARES} and the same parameter ranges as used for the training set. Importantly, the test set is completely separate from the training set and is used to determine the accuracy of the trained {\tt globalemu} network.

Using the 24,000-signal {\tt ARES} training set, we train five {\tt globalemu} networks each with a different network architecture. We test a similar, although less comprehensive, grid of architectures as those tested in \citet{globalemu} (see their Figure 8): [8, 8, 8], [64, 64], [16, 16, 16, 16], [16, 16, 16], [32, 32, 32]; where the values of each component in a given bracket are the numbers of nodes in each hidden layer, and the number of components in each bracket is the number of layers. The network stops learning once the loss function does not improve by $10^{-5}$ within the last twenty epochs of training, which ensures the trained network is as accurate as possible for the chosen network architecture.\footnote{Training a network until the loss function on the training set stops decreasing can sometimes result in overfitting, where the network learns the training set specifically rather than the underlying patterns. We test for overfitting in our network trained with an architecture of 3 hidden layers of 32 nodes by comparing the distribution of loss values (i.e., RMSE normalized by the maximum $|\delta T_b|$) across the training and test data sets (see also \citealt{globalemu}). We find that the loss distributions for the test and training sets, when emulated with the trained neural network, are nearly identical, and so we can consequently conclude that the network is not overfitting the training data.} For the data pre-processing step that is required before training the network (see section 4 of \citealt{globalemu}), we turn off the astrophysics-free-baseline (AFB) subtraction and resampling options because we find that they have a slightly negative impact on the accuracy of the resulting trained network. The lack of benefit from the pre-processing steps may be due to the fact that the `astrophysics free' Dark Ages comprises a small portion of our simulated signals.

We determine the accuracy of each trained network by evaluating them at the parameter values of the 2,000 {\tt ARES} signals in the test set and comparing the resulting emulated signals to their corresponding `true' signals. The top panel of Figure~\ref{fig:testset} shows a subset of the test set (in black) plotted along with the corresponding emulations (in red) generated by the {\tt globalemu} network used for analyses presented in this work. The bottom panel of Figure~\ref{fig:testset} shows the residuals between the emulated and `true' signals in the test set. We find that the network architecture of [32, 32, 32] gives the lowest mean RMSE of 1.25 mK (with a maximum RMSE of $18.5$ mK) between the 2,000 emulated and `true' signals (see the horizontal dotted, red line in the bottom panel of Figure~\ref{fig:testset}), while the other network architectures gave mean RMSEs ranging from 1.8 mK to 4.5 mK. Network training for the architecture [32, 32, 32] took 10 hours, as performed on a 2018 MacBook Pro with a 6-core i9 processor and 32 GB of memory. The mean RMSE of 1.25 mK is comparable to or better than those achieved in other studies that trained {\tt globalemu} on large training sets (e.g., \citealt{Bevins22a, Bevins22b, Bevins23}), and \citet{globalemu} also found [32, 32, 32] to give the lowest mean RMSE of the trained network.

The efforts described above to optimize the accuracy of the {\tt ARES}-trained {\tt globalemu} network provide robustness to the accuracy limits determined in Section~\ref{sec:results}. Even so, the small RMSE of the trained network should contribute to bias on the resulting emulated parameter constraints. We briefly investigate this by determining whether or not there is a correlation in the test set between the depth of a signal's Cosmic Dawn (CD) trough and the accuracy of its corresponding emulation. We find no statistically significant correlation between CD trough depth and the mean, median, or maximum emulation residual, obtaining Kendall rank and Pearson correlation coefficients between $-0.1$ and $-0.6$ with p-values all $<10^{-3}$. Therefore, we infer that emulated posterior biases are not correlated to emulation residuals in the CD trough depth, although we defer to future work a detailed investigation of the relationship between RMSE network uncertainties and the accuracy of the emulator model constraints.

\subsection{Mock Data}
\label{subsec:mockdata}

For all analyses, we fit the same mock data realization for the global 21-cm signal and galaxy UVLF at $z=5.9$ generated by {\tt ARES} using a fiducial set of parameter values, $\theta_0$ (see Figure~\ref{fig:trainingset} and Table~\ref{tab:params}). The fiducial values used for the four parameters that the UVLF is sensitive to (i.e., the four SFE parameters: $f_{\rm \star,0}$, $M_{\rm p}$, $\gamma_{\rm lo}$, and $\gamma_{\rm hi}$; see Equation~\ref{eqn:SFE}) were determined empirically via calibration to the $z=5.9$ UVLF measured by \citet{Bouwens15} (see \citealt{Mirocha17} for details on this calibration). For the other four astrophysical parameters that we constrain (i.e., the four `non-SFE' parameters: $c_X$, $f_{\rm esc}$, $T_{\rm min}$, and $\log N_{\rm H \RomanNumeralCaps{1}}$) we use typical, physically-motivated fiducial values based on observations or simulations.

Because the non-SFE parameters have no effect on the {\tt ARES} UVLF model, their values are not constrained by the UVLF calibration procedure. In particular, the fiducial value for $c_X$ is motivated by studies of low-$z$ star-forming galaxies (e.g., \citealt{Mineo12}), and the fiducial value for $\log N_{\rm H \RomanNumeralCaps{1}}$ is motivated by simulations (e.g., \citealt{Das17}). The difference in our fiducial values for $f_{\rm esc}$ and $\log N_{\rm H \RomanNumeralCaps{1}}$ compared to those used in \citet{Mirocha17} result in our fiducial mock global 21-cm signal (see the blue curve in Figure~\ref{fig:trainingset}) having a Cosmic Dawn trough that is located at the same frequency but is $\approx 50$ mK deeper.

The fiducial mock global 21-cm signal is created in the same manner as the training set (i.e., $z = 6-55$ with step $\delta z=0.1$), and the mock galaxy UVLF is created at the same ten magnitudes as the UVLF at $z=5.9$ measured by \citet{Bouwens15}. Therefore, the mock UVLF that we fit is a collection of ten data points that resembles the actual $z=5.9$ UVLF measured by \citet{Bouwens15}, but with small vertical offsets from the real data points due to the UVLF calibration procedure that allows us to identify the input model parameters (see the left panel of Fig. 2 in \citet{Mirocha17} for a comparison of the fiducial {\tt ARES} UVLF model and the \citet{Bouwens15} UVLF).

The noise that we add to the fiducial mock 21-cm signal is Gaussian-distributed with a standard deviation noise estimate $\sigma_{21}$. For our analyses, we test five different 21-cm noise levels \cite[including the optimistic, fiducial, and pessimistic scenarios used for the REACH radiometer in][]{REACH}: $\sigma_{21} =$ 5 mK or 10 mK (referred to as `optimistic'), $\sigma_{21} =$ 25 mK or 50 mK (referred to as `standard'), and $\sigma_{21} =$ 250 mK (referred to as `pessimistic'). We also note that the noise added to $\boldsymbol{D_{21}}$ is constant in frequency space, whereas in practice, the noise on the measured global 21-cm signal is expected to decrease with increasing frequency according to the radiometer equation. It has been suggested that such frequency dependence has little impact on the derived parameter constraints \citep{Bevins22b}, but full treatment is left for future work. For the UVLF, we use the error reported for the $z=5.9$ UVLF data from \citet{Bouwens15}.

\section{Results}
\label{sec:results}

In this section, we present the results of fitting mock global 21-cm signal data, with and without mock high-$z$ galaxy UVLF data, using various noise levels to be expected from 21-cm experiments. For the astrophysical modelling, we employ either an {\tt ARES}-trained {\tt globalemu} network or the full {\tt ARES} model.

We first discuss the posteriors obtained when jointly-fitting the mock 21-cm and UVLF data (Section~\ref{subsec:joint}, \Cref{fig:multinest_joint_25mK,fig:1Dposteriors,fig:bias}, followed by those obtained when separately fitting the individual data sets (Section~\ref{subsec:inividualfits}, Figure~\ref{fig:multinest_onlysignal_onlyUVLF}), and lastly we discuss the concept of posterior consistency in our results (Section~\ref{subsec:consistency}). Because the joint-fits produce unimodal posteriors with well-behaved means, we focus primarily on the posteriors from joint-fits when comparing {\tt globalemu} and {\tt ARES}.

We determine the accuracy of the {\tt ARES}-trained {\tt globalemu} model by comparing the mean (see top panel of Figure~\ref{fig:bias}) or the shape (see Appendix~\ref{sec:KS}) of the emulated posterior distributions to those of the `true,' full {\tt ARES} posteriors. Note that this comparison is driven by the global 21-cm signal since {\tt globalemu} does not emulate the UVLF, which we continue to model with {\tt ARES}. To our knowledge, {the recently released \sc 21cmEMU} \citep{Breitman23} is the only publicly available emulator that includes the UVLF; see, however, \cite{Kern17} for a more general emulator.

For most fits, we find necessary to use more than the default number of initial live points in {\tt MultiNest} of $n_{\rm live} = 400$ in order to fully sample the posterior and obtain convergence (see Table~\ref{tab:results} for details on fits performed). For the choice of sampling efficiency (i.e., the ratio of points accepted to points sampled), we use the recommended value for parameter estimation in {\tt MultiNest}, $e=0.8$, and for the evidence tolerance, the recommended, default value of ${\tt tol} =0.5$.\footnote{Note that when testing the decrease of this value to 0.1 it only extended the length of the run without significantly altering the final results.}~All of the triangle plots shown in this paper were generated using the Python module {\tt corner.py} (\citealt{corner}, v2.0.0, Zenodo, \href{https://doi.org/10.5281/zenodo.53155}{doi:10.5281/zenodo.53155}) with 100 bins and a Gaussian smoothing kernel of 2$\sigma$. For case examples, we tested that increasing the number of bins did not affect the essence of the results presented. The resulting posteriors were plotted using the samples and weights output by the converged nested sampling runs.

\subsection{Jointly-fitting 21-cm and UVLF Mock Data}
\label{subsec:joint}

\begin{figure*}[t]
    \includegraphics[width=\textwidth]{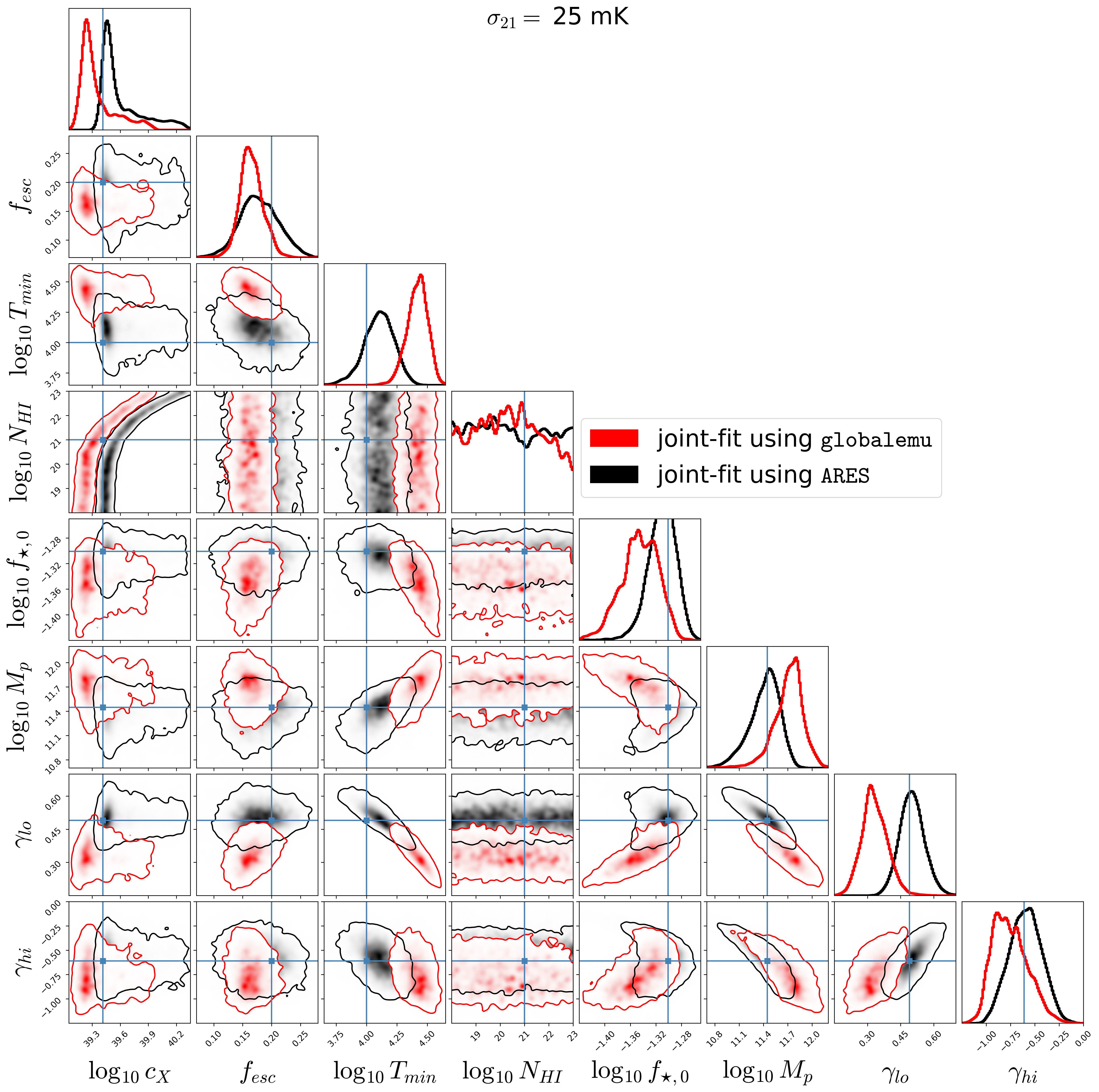}
    \caption{Marginalized 1D and 2D posterior distributions for eight astrophysical parameters in {\tt ARES} when jointly-fitting mock global 21-cm signal and UVLF data. These eight parameters control the SFE and the UV and X-ray photon production in galaxies (see Table~\ref{tab:params}). The red posterior is obtained using the {\tt ARES}-trained {\tt globalemu} network model, and the black posterior is obtained using the full {\tt ARES} model. Blue vertical and horizontal lines indicate the input, or fiducial, parameter values used to generate the mock data being fit (see Table~\ref{tab:params}), which are calibrated to real observations of the UVLF (see Section~\ref{subsec:mockdata}). The statistical noise in the 21-cm data being fit is $\boldsymbol{\sigma_{21}}=$ \textbf{25 mK}, which among the five tested we find gives the most accurate `true' {\tt ARES} posteriors with respect to the fiducial parameter values, and also highlights for which parameters {\tt globalemu} obtains biased constraints (see also Figure~\ref{fig:bias}). The UVLF data noise is the same as the error on the $z=5.9$ UVLF measurements from \citet{Bouwens15}. Contour lines in the 2D histograms represent the 95\% confidence levels, and density colormaps are shown. Axis ranges are zoomed-in with respect to the full prior ranges given in Table~\ref{tab:params}. See Table~\ref{tab:results} for further details on each fit.}
    \label{fig:multinest_joint_25mK}
\end{figure*}

\begin{figure}[ht]
    \includegraphics[scale=0.75]{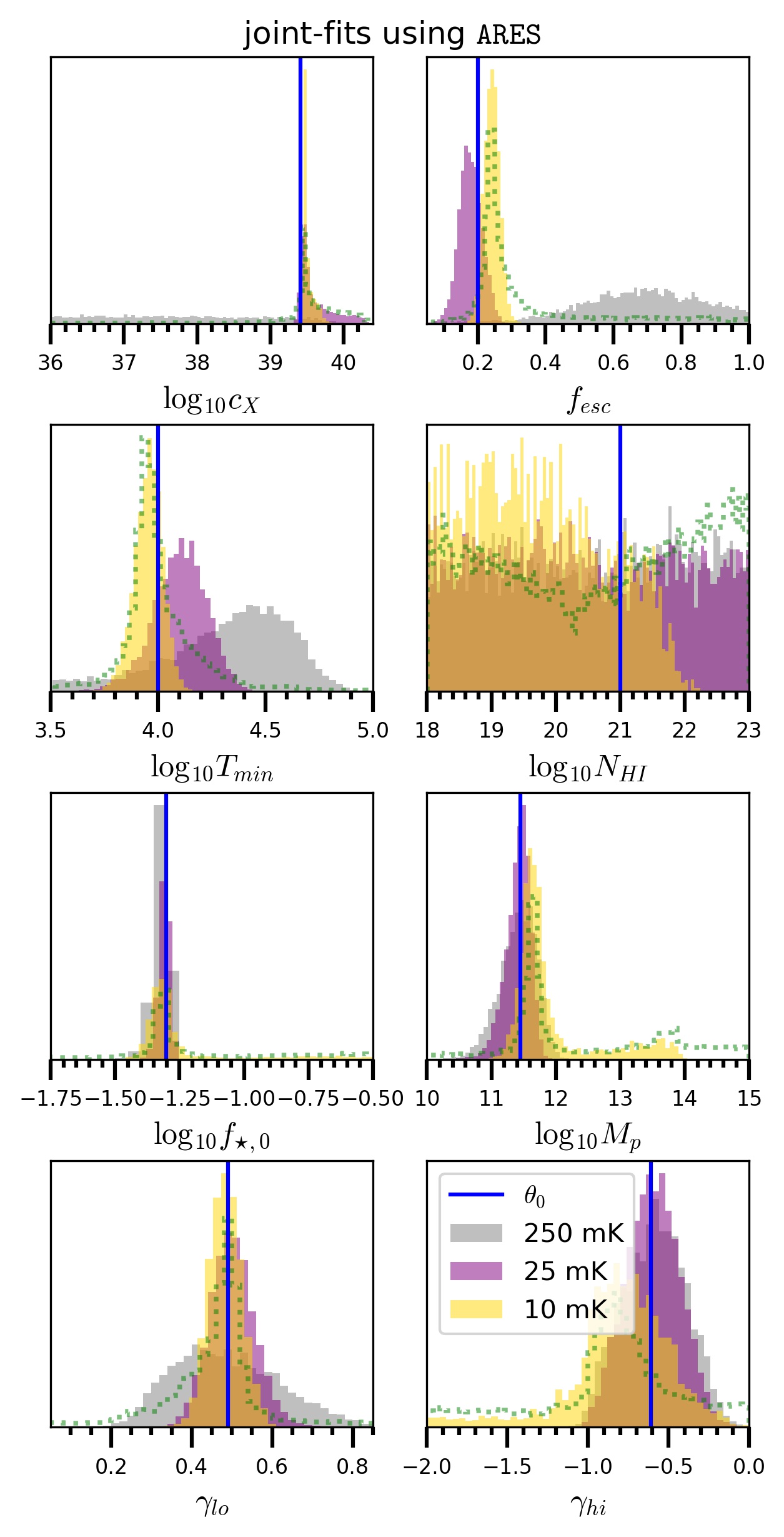}
    \caption{Marginalized 1D posterior distributions when jointly-fitting mock global 21-cm signal and UVLF data using the full {\tt ARES} model, for three different 21-cm noise levels: 10 mK (optimistic), 25 mK (standard), and 250 mK (pessimistic). These eight parameters control the SFE and the UV and X-ray photon production in galaxies (see Table~\ref{tab:params}). Blue vertical lines indicate the input, or fiducial, parameter values used to generate the mock data (see Section~\ref{subsec:mockdata}). The dotted, green histograms result from using {\tt PolyChord} with $\sigma_{21}=$ 10 mK and match well the corresponding distributions obtained by using {\tt MultiNest}. The noise on the mock UVLF being fit is the same as the error on the $z=5.9$ UVLF measurements from \citet{Bouwens15}, except for the posteriors for 10 mK shown here, for which we used twice the UVLF error to allow for a reasonable convergence time of the {\tt PolyChord} run (see Section~\ref{subsec:joint}). The posteriors for 25 mK and 250 mK are the same as those in~\Cref{fig:multinest_joint_25mK,fig:multinest_joint_250mK}, respectively. Axis ranges are zoomed-in from the full prior ranges given in Table~\ref{tab:params}.}
    \label{fig:1Dposteriors}
\end{figure}

\begin{figure}[ht]
    \includegraphics[scale=0.54]{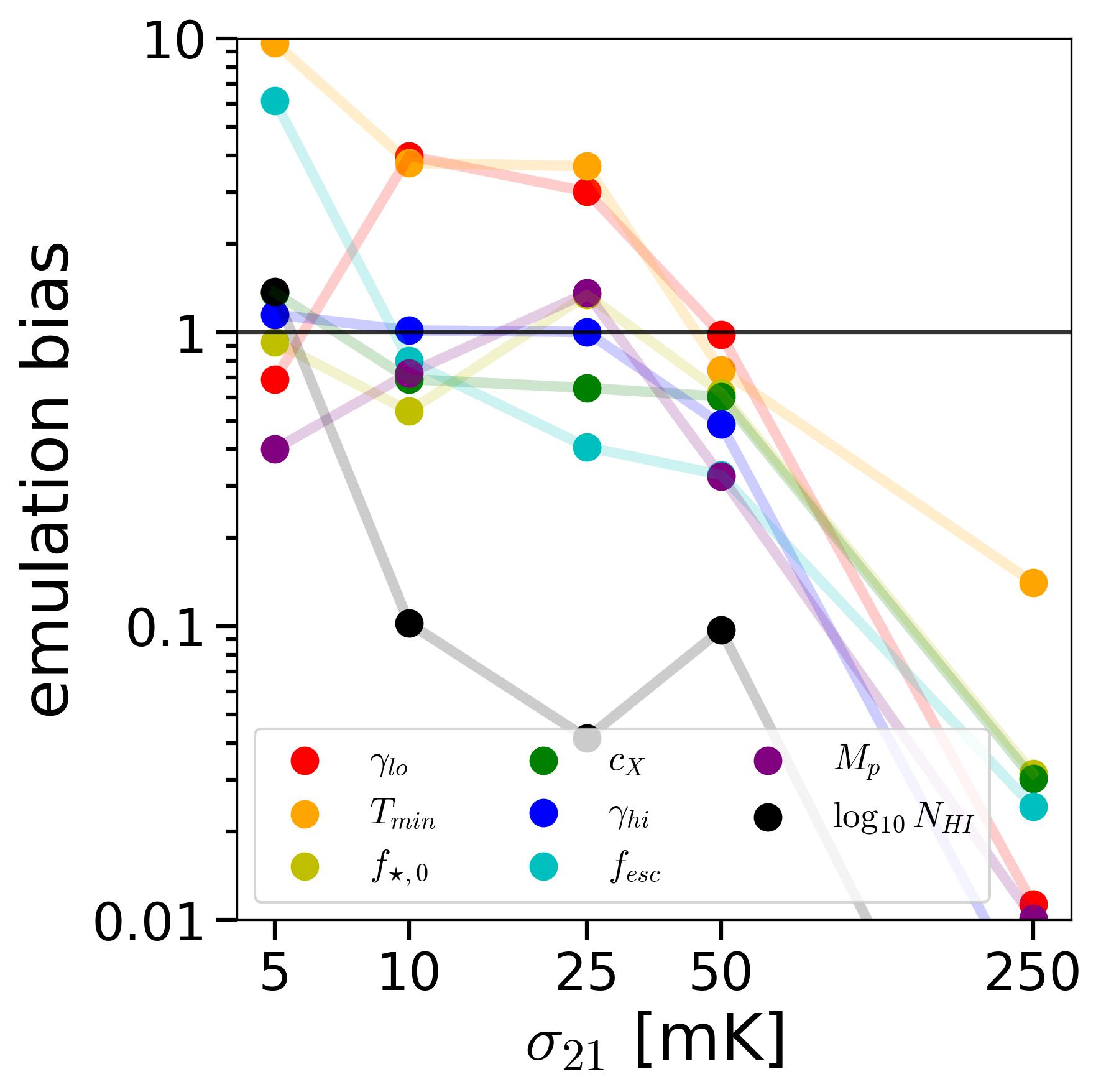} %scale=0.56
    \includegraphics[scale=0.53]{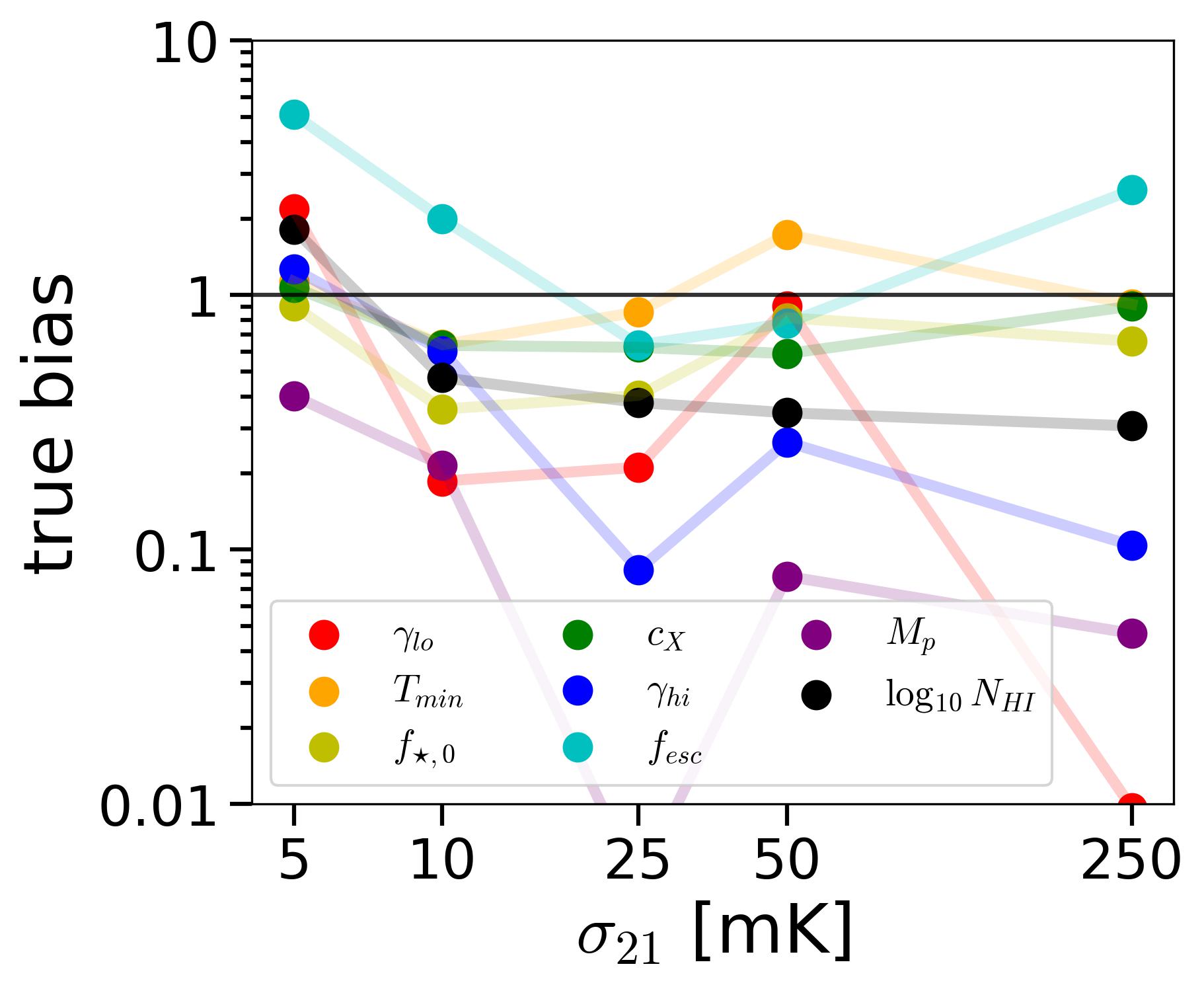}
    \caption{{\it Top:} Emulation bias (number of standard deviations, see Equation~\ref{eqn:emulationbias}) between {\tt globalemu} and {\tt ARES} for different noise levels of the mock 21-cm data being jointly-fit with the mock UVLF data.~Generally, the emulation bias decreases as the 21-cm noise level increases. For $\sigma_{21}=$ 50 mK and 250 mK, the emulation biases are $<1$ for all eight parameters, as indicated by the horizontal black line. The emulation biases for $\gamma_{\rm lo}$, $T_{\rm min}$, and $f_{\rm esc}$ can be significantly higher than the rest for certain lower 21-cm noise levels. {\it Bottom:} True bias (Equation~\ref{eqn:truebias}) between {\tt ARES} and the fiducial parameter values, for the same joint-fits. True bias is lowest at 25 mK ($<1$ for all parameters), and increases at high and low 21-cm noise levels due to increased uncertainty and difficulty in sampling, respectively (see Section~\ref{subsec:joint}). As also discussed in the text, note that the high emulation bias on $f_{\rm esc}$ at 5 mK is dominated by its high true bias.}
    \label{fig:bias}
\end{figure}

In Figure~\ref{fig:multinest_joint_25mK}, we present the posteriors obtained from a joint-fit using either {\tt globalemu} or {\tt ARES}, for a `standard' noise level of $\sigma_{21}= 25$ mK in the 21-cm data being fit. We present this as the main joint-fit result because we find that $\sigma_{21}= 25$ mK gives the least biased mean parameter values for {\tt ARES} with respect to the fiducial ones and therefore provides the best representation of the accuracy limits of the {\tt globalemu} model with respect to {\tt ARES}. In Figure~\ref{fig:1Dposteriors}, we compare the 1D posteriors obtained from joint-fits using {\tt ARES} for three characteristic 21-cm noise levels. In Figure~\ref{fig:bias}, we summarize the biases between the emulated and `true' posteriors, as well as between the `true' posteriors and the fiducial values, for the five tested 21-cm noise levels. We present the full posterior distributions obtained from joint-fits for $\sigma_{21}= 50$ mK and 250 mK in Appendix~\ref{sec:extra}.

As to be expected, because the four SFE parameters ($f_{\rm \star,0}$, $M_{\rm p}$, $\gamma_{\rm lo}$, and $\gamma_{\rm hi}$) directly determine the UVLF model (Section~\ref{subsec:model}), their posteriors are well-constrained when adding the UVLF to the 21-cm data. For joint-fits, the four SFE posteriors are unimodal and centered on the fiducial value, which is not the case when fitting only the 21-cm data as we discuss in Section~\ref{subsec:inividualfits}. Interestingly, the bimodalities in the 1D posteriors for $M_{\rm p}$ and $\gamma_{\rm lo}$ when fitting only the 21-cm data disappear when adding the UVLF data in the joint-fit, showing that the combination of both data sets can break degeneracies in the {\tt ARES} parameter space and reduce biases. 

Comparing the emulated distributions (in red) and `true' distributions (in black) in Figure~\ref{fig:multinest_joint_25mK}, we see that the {\tt globalemu} model produces similar posteriors as the {\tt ARES} model, both in shape (see Appendix~\ref{sec:KS}) and mean (top panel of Figure~\ref{fig:bias}), except for a few exceptions discussed below. In Figure~\ref{fig:bias}, we summarize the two different types of parameter biases discussed, for the different 21-cm noise levels tested: emulation bias (Equation~\ref{eqn:emulationbias}) and true bias (Equation~\ref{eqn:truebias}). Emulation bias refers to the accuracy of the emulated posterior parameter means, $\mu_{\tt globalemu}$, with respect to the `true' posterior parameter means, $\mu_{\tt ARES}$, and true bias refers to the accuracy of $\mu_{\tt ARES}$ with respect to the fiducial parameter value, $\theta_0$. We note that these two biases provide all of the information necessary to evaluate the accuracy of {\tt globalemu} and {\tt ARES}, and that defining a third bias between $\mu_{\tt globalemu}$ and $\theta_0$ does not further aid our results.

We therefore define and compute an emulation bias as the difference in the emulated and `true' posterior parameter means divided by the standard deviation of the `true' posterior, $\sigma_{\tt ARES}$:

\vspace{-4mm}
\begin{equation}
\label{eqn:emulationbias}
{\rm emulation\,bias} \equiv \frac{|\mu_{\tt globalemu} - \mu_{\tt ARES}|}{\sigma_{\tt ARES}}.
\end{equation}

\noindent In the same manner, we define a true bias between an {\tt ARES} posterior parameter mean and its fiducial value:

\begin{equation}
\label{eqn:truebias}
{\rm true\,bias} \equiv \frac{|\mu_{\tt ARES} - \theta_0|}{\sigma_{\tt ARES}}.
\end{equation}

We find that, in general, the emulation bias decreases as the 21-cm noise level increases. For $\sigma_{21}=$ 50 mK and 250 mK, all parameters' emulation biases are $\leq1$ (marked with a black, horizontal line in Figure~\ref{fig:bias}), while at lower noise levels ($\sigma_{21}=$ 5 mK, 10 mK, and 25 mK) the emulation bias raises above $1$ for certain parameters. For $\sigma_{21}=$ 10 and 25 mK, $T_{\rm min}$ and $\gamma_{\rm lo}$ have emulation biases of $3-4$, and for $\sigma_{21}=$ 5 mK, $T_{\rm min}$ and $f_{\rm esc}$ have even higher emulation biases of $\approx6-10$, while the emulation bias of $\gamma_{\rm lo}$ drops below 1.

The relatively high emulation biases on $T_{\rm min}$ and $\gamma_{\rm lo}$ are due only to the {\tt globalemu} posteriors being less accurate, given that the true biases on $T_{\rm min}$ and $\gamma_{\rm lo}$ are low (see the bottom panel of Figure~\ref{fig:bias}). In contrast, the high emulation bias on $f_{\rm esc}$ at 5 mK is influenced by the high true bias on the {\tt ARES} posterior for $f_{\rm esc}$. We find that true biases $\geq1$ at low 21-cm noise levels for $f_{\rm esc}$ and $T_{\rm min}$ also exist when using other samplers such as {\tt PolyChord} (see Figure~\ref{fig:1Dposteriors}) and {\tt emcee} \citep{emcee}, and so we infer that these biases could be due to accuracy limitations of the sampling algorithms to produce unbiased constraints at very low noise levels. Future work could further explore sampling biases at such low noise levels by using other algorithms such as {\tt dynesty} \citep{Speagle20} and in particular {\tt UltraNest} \citep{Buchner16, Buchner19, UltraNest}, which was created for the purpose of mitigating bias in complex posteriors.

We can also compare the final evidences output from the nested sampling analyses and compute the Bayes factor (i.e., the ratio of evidences, or difference of log evidences) to select the favored model given the data and priors \citep{Trotta08}. For $\sigma_{21}= 25$ mK, the Bayes factor between {\tt globalemu} and {\tt ARES} is 0.6; for 50 mK, it is 2.7; and for 250 mK, 1.2. The natural logarithm of these Bayes factors being $<1$ %of order unity 
indicates that there is no preference for one model over the other in fitting the mock data (see e.g., \citealt{KassRaftery95, jeffreys1998, Trotta08}).

In Figure~\ref{fig:1Dposteriors}, we see that as $\sigma_{21}$ increases, the non-SFE posteriors become less constrained around the fiducial value, except for $\log N_{\rm H \RomanNumeralCaps{1}}$ which is unconstrained at all noise levels. At high/pessimistic $\sigma_{21}$, the 21-cm data provides much less constraining power, which causes degeneracies in the 8-dimensional {\tt ARES} parameter space to grow larger (i.e., the space becomes flatter). This subsequently widens the posterior distributions for those parameters that are most sensitive to the 21-cm data (see also Section~\ref{subsec:consistency}). For the 21-cm noise level of 250 mK, the true biases are $\approx1$ for $c_X$ and $T_{\rm min}$, and $\approx2.5$ for $f_{\rm esc}$. In contrast, for $\sigma_{21}= 10$ mK, 25 mK, and 50 mK, there is no true bias $\geq1$, except for $T_{\rm min}$ at 50 mK and $f_{\rm esc}$ at 10 mK, which each have true bias of $\approx2$ (see bottom panel of Figure~\ref{fig:bias}).

As briefly mentioned, we performed one joint-fit using the {\tt PolyChord} nested sampling algorithm to compare the result to an equivalent joint-fit using {\tt MultiNest}. In Figure~\ref{fig:1Dposteriors}, the posteriors from {\tt PolyChord} are shown as dotted green histograms, and the equivalent posteriors from {\tt MultiNest} are shown in solid yellow. For the 21-cm data being fit we assume the optimistic noise level $\sigma_{21}= 10$ mK, and for the UVLF we assume twice the error on the $z=5.9$ UVLF measurements from \citet{Bouwens15} (i.e., `2xB+15'). We use `2xB+15' UVLF error instead of `B+15' because this allows the {\tt PolyChord} run to converge in a more reasonable amount of time. In addition, we find that doing so has no effect on the non-SFE posteriors and only slightly increases the width of the SFE posteriors. We find close agreement between the posterior distributions and final evidences (see Table~\ref{tab:results}) obtained when using {\tt PolyChord} versus those when using {\tt MultiNest}. Comparing the two runs, we find that {\tt PolyChord} required 28 times more likelihood evaluations to reach roughly the same result (with an acceptance rate of 0.38\% versus 8.7\% for {\tt MultiNest}; see Table~\ref{tab:results}).~{\tt PolyChord}, however, is expected to become more efficient than {\tt MultiNest} for a larger number of parameters~\citep{Handley15a}, and could thus be a better choice for 21-cm analyses including additional free parameters to account for systematics such as the beam-weighted foreground, RFI, sub-surface conditions, etc.

\subsection{Fitting Individual Mock Data Sets}
\label{subsec:inividualfits}

\begin{figure*}[t]
    \includegraphics[width=\textwidth]{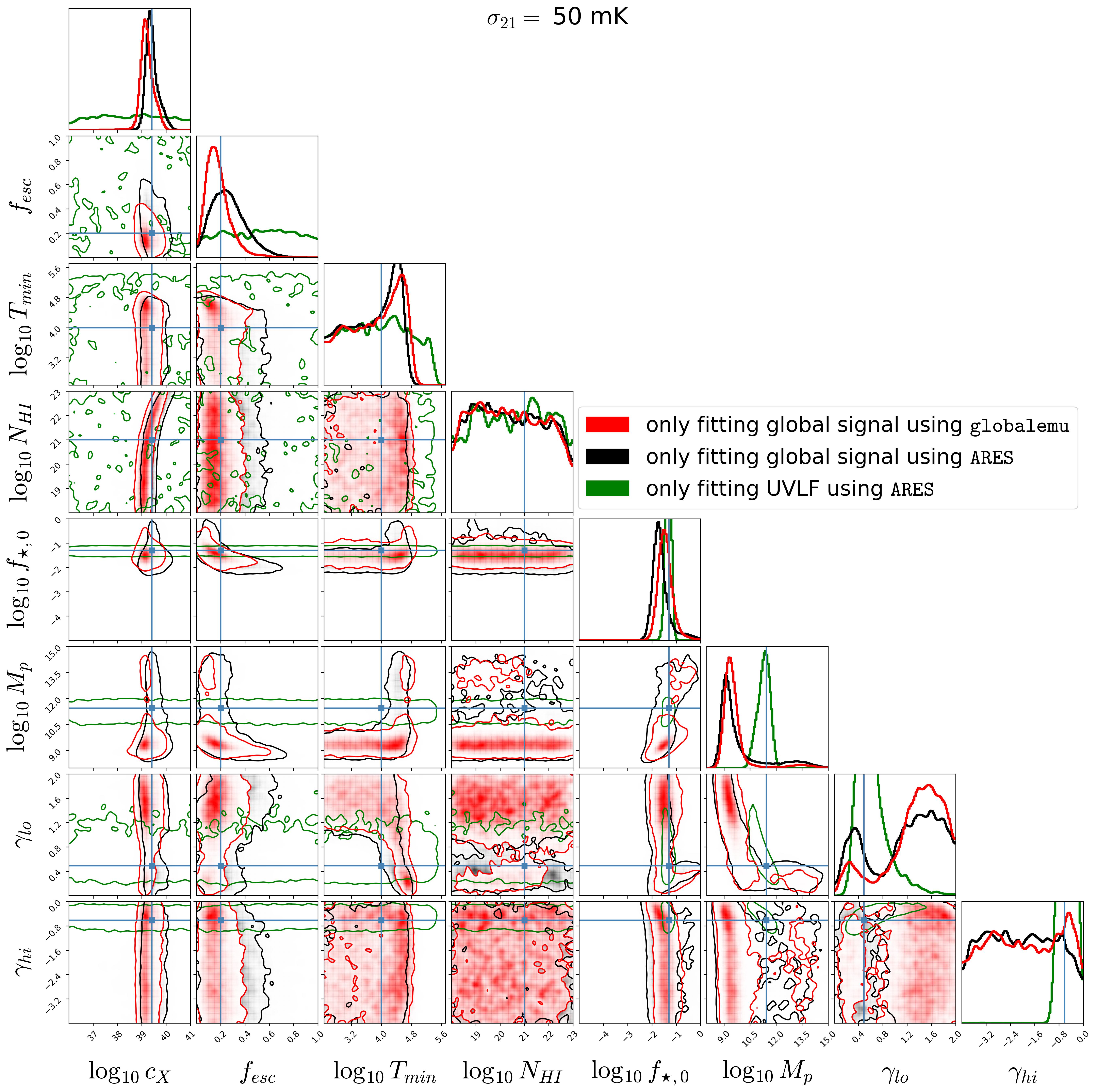}
    \caption{Marginalized 1D and 2D posterior distributions obtained when fitting either mock global 21-cm signal data (red and black) or mock UVLF data (green). All is the same as in Figure~\ref{fig:multinest_joint_25mK}, except that the statistical noise in the 21-cm data being fit is $\boldsymbol{\sigma_{21}}=$ \textbf{50 mK}, and the axis ranges are the full prior ranges given in Table~\ref{tab:params}. See Table~\ref{tab:results} for further details on each fit.}
    \label{fig:multinest_onlysignal_onlyUVLF}
\end{figure*}

In Figure~\ref{fig:multinest_onlysignal_onlyUVLF}, we present the posterior distributions when separately fitting our individual mock data sets. When fitting only the 21-cm data, using either the full {\tt ARES} model (in black) or the {\tt ARES}-trained {\tt globalemu} model (in red) for $\sigma_{21}= 50$ mK, the posterior presents large degeneracies and in general larger true biases than the corresponding joint-fit at the same $\sigma_{21}$ (shown in Figure~\ref{fig:multinest_joint_50mK}). In particular, for the SFE parameters, bimodalities and degeneracies exist when fitting only the global signal that are removed when jointly-fitting the UVLF (see Section~\ref{subsec:joint}). Among the four SFE parameters, $M_{\rm p}$ and $\gamma_{\rm hi}$ are the least constrained when fitting only the 21-cm data. This is expected because these two parameters control the brightest sources, which contribute relatively little to the global photon budget, making the global signal rather insensitive to these parameters and motivating the inclusion of the UVLF data to aid these constraints. In addition, even though the posteriors of the non-SFE parameters, $c_X$, $f_{\rm esc}$, $T_{\rm min}$, and $\log N_{\rm H \RomanNumeralCaps{1}}$, remain largely the same after adding the UVLF data, the joint-fit does significantly reduce the presence of long tails in these parameters, in particular for $f_{\rm esc}$ and $T_{\rm min}$.

When only fitting the UVLF data (green posterior in Figure~\ref{fig:multinest_onlysignal_onlyUVLF}), we find as expected strong constraints on the SFE parameters and the lack of constraints on the rest. This is because the {\tt ARES} UVLF model only depends on the four SFE parameters and is independent of the other four. The green together with the black or red posteriors in Figure~\ref{fig:multinest_onlysignal_onlyUVLF} illustrate how jointly-fitting the UVLF with the 21-cm data is expected to break significant degeneracies in this parameter space, to obtain the tight constraints shown in \Cref{fig:multinest_joint_25mK}.

Comparing the red and black constraints from the 21-cm data in Figure~\ref{fig:multinest_onlysignal_onlyUVLF}, we find that using the {\tt ARES}-trained {\tt globalemu} model produces rather similar 1D and 2D posterior distributions to those from the full {\tt ARES} model, with all emulation biases $<1$, except for $f_{\rm esc}$ which has an emulation bias of $\approx1$. As stated in Table~\ref{tab:results}, the runs using {\tt globalemu} and {\tt ARES} reach nearly the same final evidence, further demonstrating the agreement between the two results. This close agreement shows that {\tt globalemu} is able to represent the {\tt ARES} parameter space more easily when the constraints are significantly weaker with respect to those from the joint-fit with the UVLF data.

\subsection{Posterior Consistency}
\label{subsec:consistency}

Bayesian consistency of a posterior distribution is the concept that as the number of data observations grows, the posterior distribution converges on the truth \citep{Schwartz1965OnBP}\footnote{We also refer the reader to Prof. Surya Tapas Tokdar’s notes on Bayesian consistency: \url{http://www2.stat.duke.edu/~st118/sta941/Asymp.pdf}}. A posterior is considered consistent if it eventually concentrates on the true parameter value as the number of degrees of freedom in the data vector increases to infinity. As shown in Figure~\ref{fig:1Dposteriors}, we observe posterior consistency when comparing the 1D posteriors obtained for decreasing levels of the 21-cm noise: larger integration times result in posteriors generally becoming more peaked around the input, fiducial values (marked by blue lines). As briefly mentioned in Section~\ref{subsec:joint}, for lower integration times (i.e., higher $\sigma_{21}$), the 21-cm data provides relatively little constraining power, which grows the covariance in the multi-dimensional parameter space, producing probability density biases\footnote{Additionally, this should produce further departures from the standard assumptions taken in building the likelihood in Equation~\ref{eqn:likelihood} \citep[see e.g.,][]{PrelogovicMesinger23}, and likely contribute to bias the posterior.}. As expected from Bayesian consistency, we thus find that the posteriors are more biased from their fiducial values at increasing noise levels.

Posterior consistency is most apparent for these four parameters: $c_X$, $f_{\rm esc}$, $T_{\rm min}$, and $\gamma_{\rm lo}$. Their pessimistic noise level posteriors ($\sigma_{21}=$ 250 mK; gray in Figure~\ref{fig:1Dposteriors}) are clearly not centered on their fiducial values, presenting a relatively slow `rate of convergence,' while the three SFE parameters $f_{\rm \star,0}$, $M_{\rm p}$, and $\gamma_{\rm hi}$ have faster rates of convergence and thus require less integration time to concentrate on their input, fiducial values.~As also shown in the triangle plots above, $\log N_{\rm H \RomanNumeralCaps{1}}$ remains largely unconstrained for all the noise levels, though {\tt globalemu} still accurately emulates its posterior.

\begin{table*}[t]
    \setlength{\tabcolsep}{0.25em}
    \def\arraystretch{0.95}
    \caption{Summary of key nested sampling analyses}
    \begin{center}
    \begin{tabular}{ccccccccc}
    \toprule
    Type of mock data being fit & Model used in likelihood & $\sigma_{21}$ & $\sigma_{\rm UVLF}$ & $n_{\rm live}$ & $\log {\it Z}$ & $n_{\rm evaluations}$ & $f_{\rm accept}$ & sec./eval.\\
    & & (mK) & (${\rm mag}^{-1} {\rm cMpc}^{-3}$) & & & & & (s) \\
    \hline
    \multirow{9}{*}{both UVLF and global 21-cm signal} & \multirow{3}{*}{\tt globalemu} & 25 & B+15 & 600 & -280.6 $\pm$ 0.2 & 94,843 & 0.147 & 24.60\\
    & & 50 & B+15 & 600 & -278.5 $\pm$ 0.2 & 91,614 & 0.163 & 24.35\\
    & & 250 & B+15 & 600 & -274.4 $\pm$ 0.1 & 44,628 & 0.244 & 24.20\\
    \cline{2-9}
    & \multirow{5}{*}{\tt ARES} & 5 & B+15 & 800 & -290.1 $\pm$ 0.3 & 445,729 & 0.052 & 34.20\\
    & & 10 & 2xB+15 & 400 & -282.2 $\pm$ 0.2 & 128,745 & 0.087 & 35.64\\
    & & 25 & B+15 & 800 & -280.0 $\pm$ 0.1 & 129,367 & 0.154 & 36.04\\
    & & 50 & B+15 & 800 & -275.8 $\pm$ 0.1 & 104,890 & 0.185 & 34.62\\
    & & 250 & B+15 & 800 & -273.2 $\pm$ 0.1 & 57,448 & 0.254 & 37.56\\
    \cline{2-9}
    & \multirow{1}{*}{{\tt ARES} using {\tt PolyChord}} & 10 & 2xB+15 & 400 & -284.3 $\pm$ 0.2 & 3,650,406 & 0.004 & 39.92\\
    \hline  
    \multirow{4}{*}{only global 21-cm signal} & \multirow{2}{*}{\tt globalemu} & 25 & ... & 1,200 & -268.9 $\pm$ 0.1 & 142,805 & 0.145 & 0.01\\
    & & 50 & ... & 1,200 & -266.5 $\pm$ 0.1 & 86,308 & 0.202 & 0.01\\
    \cline{2-9}
    & \multirow{2}{*}{\tt ARES} & 25 & ... & 1,200 & -268.8 $\pm$ 0.1 & 166,331 & 0.128 & 17.23\\
    & & 50 & ... & 1,200 & -266.2 $\pm$ 0.1 & 103,231 & 0.176 & 18.11\\
    \hline
    \multirow{1}{*}{only UVLF} & \multirow{1}{*}{\tt ARES} & ... & B+15 & 400 & -11.6 $\pm$ 0.2 & 15,578 & 0.348 & 16.37\\
    \bottomrule
    \end{tabular}
    \begin{tablenotes}
    \small
    \item The information provided for each fit are the noise level of the mock 21-cm signal ($\sigma_{21}$) and/or UVLF ($\sigma_{\rm UVLF}$) being fit, the number of initial live points used ($n_{\rm live}$), and the final output metrics, including the evidence ($\log {\it Z}$), the total number of likelihood evaluations ($n_{\rm evaluations}$), the acceptance rate ($f_{\rm accept}$), and average CPU-time required per evaluation (sec./eval.). `B+15' denotes that the UVLF error used is the same as that of the $z=5.9$ UVLF data by \citet{Bouwens15} (see Section~\ref{subsec:mockdata}). All fits shown were performed using {\tt MultiNest}, except for one joint-fit for which we used {\tt PolyChord}, the result of which is consistent with the equivalent {\tt MultiNest} fit (see Figure~\ref{fig:1Dposteriors}). The fit using {\tt PolyChord} required over an order of magnitude more computational time to converge compared to the equivalent {\tt MultiNest} fit, and so we used twice the `B+15' UVLF error to aid convergence in a reasonable amount of time without significantly affecting the results (see Section~\ref{subsec:joint}). The results from each fit included here are presented in Section~\ref{sec:results} and Appendix~\ref{sec:extra} (see~\Cref{fig:multinest_joint_25mK,fig:1Dposteriors,fig:bias,fig:multinest_onlysignal_onlyUVLF} and~\Cref{fig:multinest_joint_50mK,fig:multinest_joint_250mK}), except for the $\sigma_{21}=$ 25 mK only global signal fits.
    \end{tablenotes}
    \end{center}
    \label{tab:results}
\end{table*}

\section{Conclusions}
\label{sec:conclusions}

In this paper, we present the 1D and 2D posterior distributions for eight astrophysical parameters in {\tt ARES} obtained when fitting mock data of the global 21-cm signal and/or the high-$z$ galaxy UVLF via nested sampling. We compare for the first time the posteriors obtained from a global 21-cm signal emulator to those obtained using the full model on which it is trained, at various 21-cm noise levels. Use of an emulator such as {\tt globalemu} is desirable as it speeds up model evaluations by several orders of magnitude, but the accuracy of such constraints is poorly understood. The eight parameters employed control in {\tt ARES} the star formation efficiency (SFE) and the efficiency of UV and X-ray photon production per unit star formation in galaxies (see Table~\ref{tab:params}).

We assess the accuracy of the parameter constraints obtained by an {\tt ARES}-trained {\tt globalemu} network and determine for which parameters and 21-cm noise levels {\tt globalemu} is biased compared to {\tt ARES}. We test optimistic, standard, and pessimistic 21-cm noise levels ranging between $\sigma_{21} =$ 5 mK and 250 mK to show the astrophysical constraints that can be expected for non-systematics-limited 21-cm experiments. We optimize the accuracy of the trained {\tt globalemu} network by testing multiple network architectures and training set sizes, obtaining a mean RMSE between the emulated and true {\tt ARES} signals in the test set of 1.25 mK.

We find that adding the UVLF to the 21-cm data provides significant improvements to the constraints on the four SFE parameters, and it has little to no effect on the constraints on the non-SFE parameters. These results imply that combining 21-cm observations with {\it HST} and {\it JWST} measurements of the UVLF at different redshifts may provide key insights into the suggested redshift evolution of the star formation efficiency and the degree of stochasticity.

The {\tt ARES}-trained {\tt globalemu} model produces relatively accurate posteriors with respect to the `true' {\tt ARES} model at the tested 21-cm noise levels, both in shape and mean, except for the following. In particular, $T_{\rm min}$ and $\gamma_{\rm lo}$ present significant emulation biases at $\sigma_{21}=$ 25 mK or lower, for which {\tt globalemu} overpredicts $T_{\rm min}$ and underpredicts $\gamma_{\rm lo}$ by $\approx 3-4\sigma$ (see the top panel of Figure~\ref{fig:bias}, and Figure~\ref{fig:multinest_joint_25mK} for the full posterior distributions), except for at $\sigma_{21}=$ 5 mK, where $\gamma_{\rm lo}$ has a negligible bias. For noise levels of $\sigma_{21}=$ 50 mK and 250 mK, the {\tt globalemu} emulator reproduces the posterior means found by {\tt ARES} at the 68\% confidence level for all eight parameters (see the top panel of Figure~\ref{fig:bias}, and Appendix~\ref{sec:extra} for the full posterior distributions).

When examining the 1D posteriors obtained from joint-fits at various noise levels in Figure~\ref{fig:1Dposteriors}, we find that as the noise in the 21-cm data decreases, the 1D posteriors become more concentrated around their input, fiducial values, as expected for `posterior consistency.' For standard noise levels of $\sigma_{21}= 25$ mK and 50 mK, the true biases for all parameters are $<1$, except for at $\sigma_{21}= 50$ mK where $T_{\rm min}$ has a true bias of $\approx1.5$. For the pessimistic noise level of $\sigma_{21}=$ 250~mK, three parameters ($c_X$, $T_{\rm min}$, and $f_{\rm esc}$) have `true' {\tt ARES} posterior means that are $\approx 1-3$ $\sigma$ away from their fiducial value (i.e., have `true biases' $\approx 1-3$; see the bottom panel of Figure~\ref{fig:bias}). This indicates a slow rate of convergence for these parameter fits and the need for a longer integration time to achieve posteriors centered around the true value.

In summary, this work provides insights on the statistical constraints that are achievable from global 21-cm measurements in combination with high-$z$ UVLF data when using an emulator. We obtain strong constraints on eight {\tt ARES} parameters when jointly-fitting such data using either the full {\tt ARES} model or an {\tt ARES}-trained {\tt globalemu} model. The most accurate {\tt ARES} constraints are achieved for a 21-cm noise level of $25$ mK, where all eight {\tt ARES} parameter means are within $1\sigma$ of their fiducial values.~At this noise level, however, {\tt globalemu} overpredicts $T_{\rm min}$ and underpredicts $\gamma_{\rm lo}$. For larger noise levels of $50$ and 250 mK, while in general the true biases increase, the emulated and true posteriors match more closely such that their parameter means are within $1\sigma$ of each other.

\begin{acknowledgments}
We thank the anonymous reviewer for their detailed comments that helped improve the manuscript. We thank Harry Bevins for useful discussions. This work was directly supported by the NASA Solar System Exploration Research Virtual Institute cooperative agreement 80ARC017M0006. This work was also partially supported by the Universities Space Research Association via D.R. using internal funds for research development. We also acknowledge support by NASA grant 80NSSC23K0013. J.M. was supported by an appointment to the NASA Postdoctoral Program at the Jet Propulsion Laboratory/California Institute of Technology, administered by Oak Ridge Associated Universities under contract with NASA. This work utilized the Blanca condo computing resource at the University of Colorado Boulder. Blanca is jointly funded by computing users and the University of Colorado Boulder.
\end{acknowledgments}

\software{This research relies heavily on the {\sc python} \citep{python} open source community, in particular, {\sc numpy} \citep{numpy}, {\sc matplotlib} \citep{matplotlib}, {\sc scipy} \citep{scipy}, and {\sc jupyter} \citep{jupyter}. This research also utilized {\tt MultiNest} \citep{Feroz08, Feroz09, Feroz19}, {\tt PolyChord} \citep{Handley15a, Handley15b}, and {\tt globalemu} \citep{globalemu}.}

\appendix 
\renewcommand\thefigure{\thesection.\arabic{figure}}
\section{K-S tests comparing emulated and `true' posterior distributions} 
\label{sec:KS}

To further evaluate the accuracy of the {\tt ARES}-trained {\tt globalemu} model, we also perform a two-sample Kolmogorov-Smirnov (K-S) test on each pair of emulated and `true' 1D posteriors obtained from joint-fits, shown in \Cref{fig:multinest_joint_50mK,fig:multinest_joint_25mK,fig:multinest_joint_250mK}. We compute the cumulative distribution functions (CDFs) for each 1D marginalized posterior PDF (48 total: 16 from $\sigma_{21}=$ 25 mK, 16 from $\sigma_{21}=$ 50 mK, and 16 from $\sigma_{21}=$ 250 mK) and employ {\tt scipy.stats} to calculate the K-S statistics and associated $p$-values for each of the 24 pairs of CDFs.

We find that the $p$-values are all $>0.05$, except for at $\sigma_{21}=$ 250 mK where two parameters ($M_{\rm p}$ and $\gamma_{\rm hi}$) have $p<0.05$. For six of the eight parameters, the $p$-values are highest at $\sigma_{21}= 50$ mK, which is the same 21-cm noise level that we found gives the most similar means of the emulated and `true' posteriors (Figure~\ref{fig:bias}). We also computed the K-S statistics between the emulated and true 1D posteriors obtained when fitting only the global signal (Figure~\ref{fig:multinest_onlysignal_onlyUVLF}) and found that $p>0.05$ for all eight parameters. Therefore, based on the K-S tests, we conclude that the null hypothesis that the emulated and `true' posteriors originate from the same parent distribution is not rejected. We note that even though {\tt globalemu} is deemed to be a good representation of {\tt ARES} based on K-S tests and the Bayes factor (see Section~\ref{subsec:joint}), significant biases exist on the emulated posteriors for $T_{\rm min}$ and $\gamma_{\rm lo}$, in particular at lower 21-cm noise levels of $\sigma_{21} \lesssim$ 25 mK (see Figure~\ref{fig:bias}).

\section{Posterior distributions from joint-fits for 21-cm noise levels of 50 mK and 250 mK}
\label{sec:extra}
\setcounter{figure}{0}

\begin{figure*}[t]
    \includegraphics[width=\textwidth]{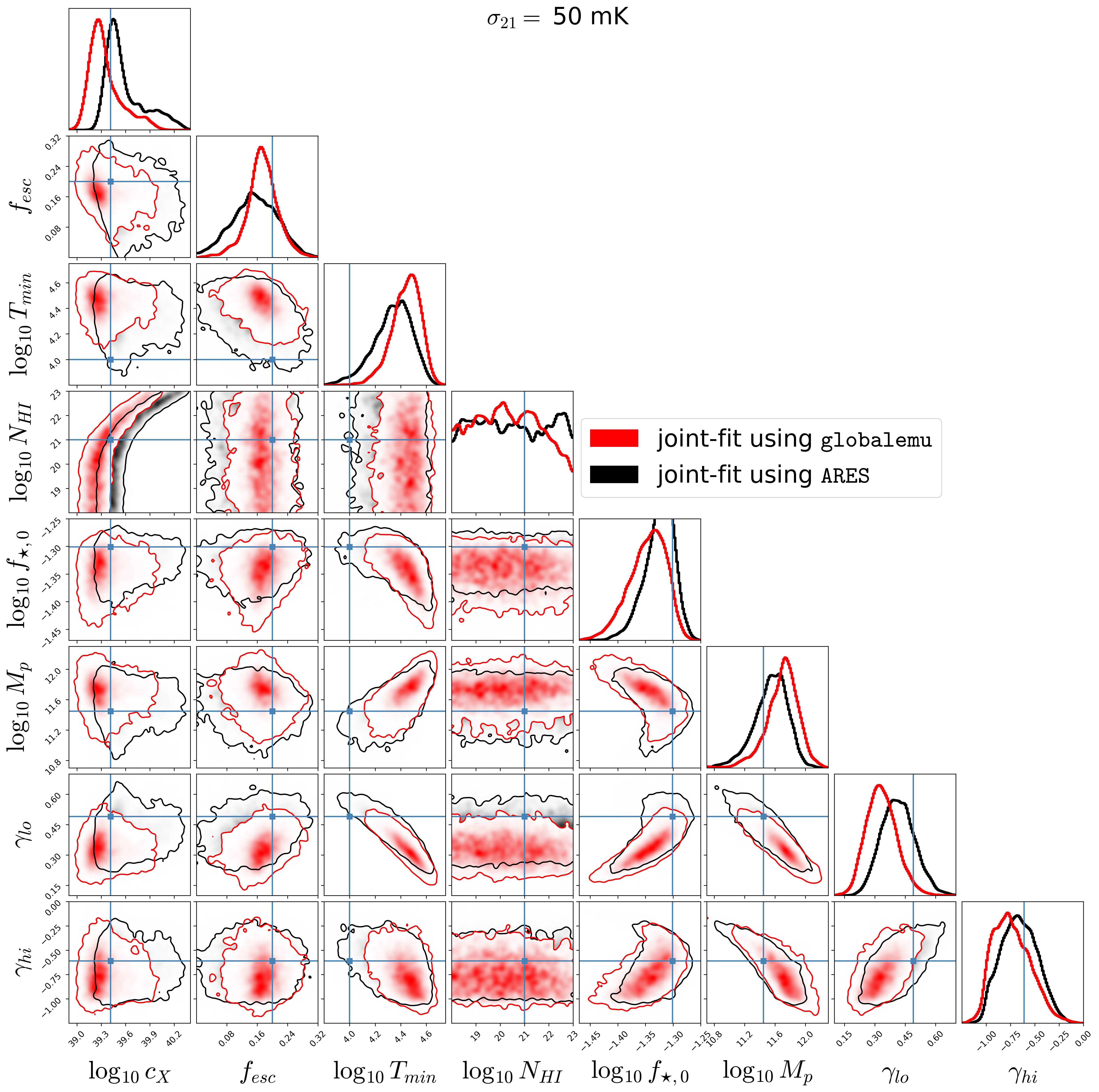}
    \caption{Marginalized 1D and 2D posterior distributions for eight astrophysical parameters in {\tt ARES} when jointly-fitting mock global 21-cm signal and UVLF data. All is the same as in \Cref{fig:multinest_joint_25mK}, except that the noise in the 21-cm data being fit is $\boldsymbol{\sigma_{21}}=$ \textbf{50 mK}.}
    \label{fig:multinest_joint_50mK}
\end{figure*}

\begin{figure*}[t]
    \includegraphics[width=\textwidth]{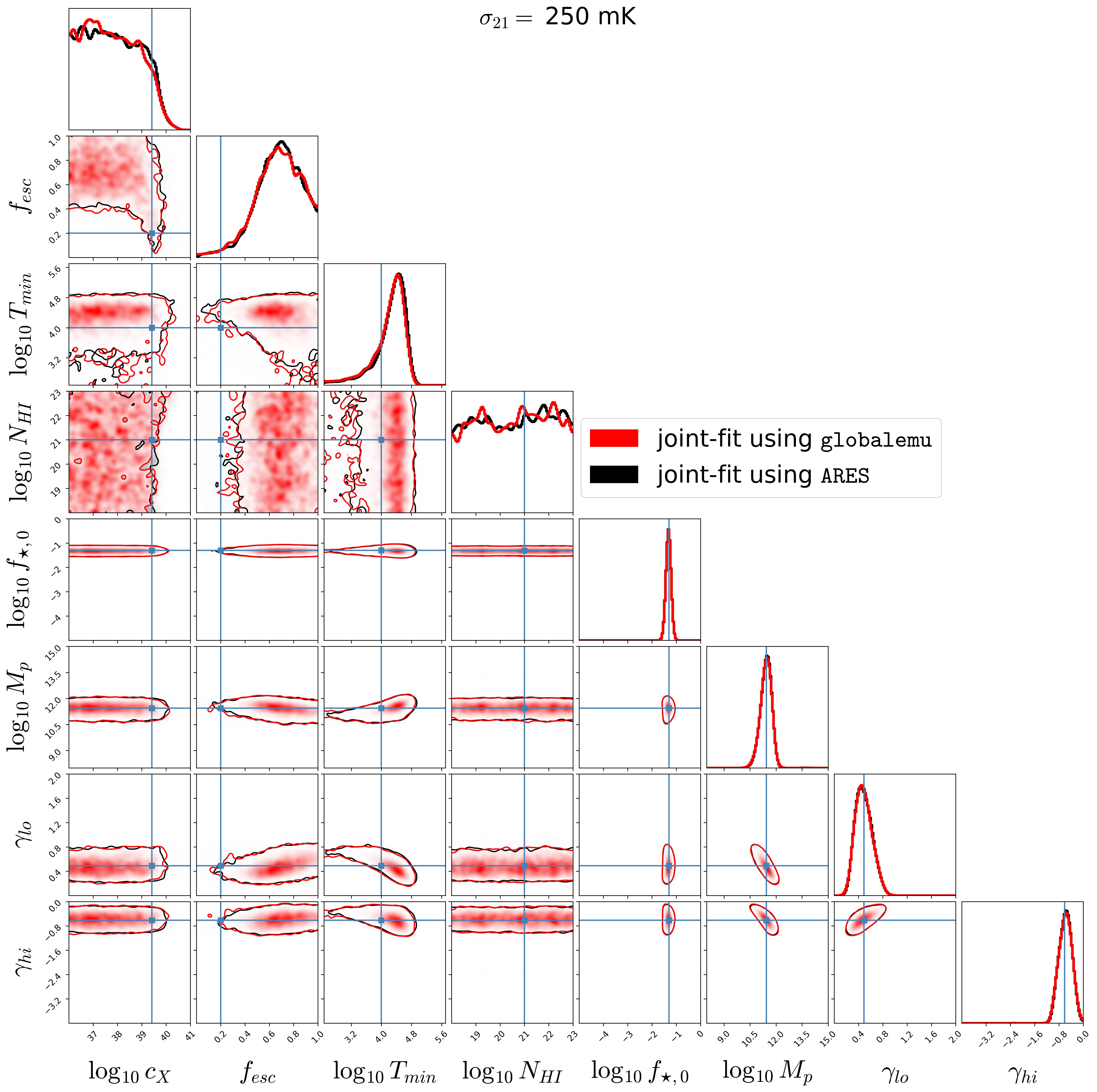}
    \caption{Marginalized 1D and 2D posterior distributions for eight astrophysical parameters in {\tt ARES} when jointly-fitting mock global 21-cm signal and UVLF data. All is the same as in \Cref{fig:multinest_joint_50mK,fig:multinest_joint_25mK}, except that the noise in the 21-cm data being fit is $\boldsymbol{\sigma_{21}}=$ \textbf{250 mK}, and the axis ranges are zoomed-out to show the full prior ranges.}
    \label{fig:multinest_joint_250mK}
\end{figure*}

In Figures~\ref{fig:multinest_joint_50mK} and~\ref{fig:multinest_joint_250mK}, we present the full posterior distributions obtained from mock 21-cm and UVLF data joint-fits for 21-cm noise levels of $\sigma_{21}=$ 50 mK and 250 mK, respectively. As discussed in Section~\ref{subsec:joint}, we found that these noise levels give the best match (i.e., most similar parameter means) between the emulated and `true' posteriors (see the top panel of Figure~\ref{fig:bias}). The 1D posteriors for $\sigma_{21}= 250$ mK are also shown in Figure~\ref{fig:1Dposteriors}.

\bibliography{dj23}{}

\begin{thebibliography}{}
\expandafter\ifx\csname natexlab\endcsname\relax\def\natexlab#1{#1}\fi
\providecommand{\url}[1]{\href{#1}{#1}}
\providecommand{\dodoi}[1]{doi:~\href{http://doi.org/#1}{\nolinkurl{#1}}}
\providecommand{\doeprint}[1]{\href{http://ascl.net/#1}{\nolinkurl{http://ascl.net/#1}}}
\providecommand{\doarXiv}[1]{\href{https://arxiv.org/abs/#1}{\nolinkurl{https://arxiv.org/abs/#1}}}

\bibitem[{{Anstey} {et~al.}(2023){Anstey}, {de Lera Acedo}, \& {Handley}}]{Anstey23}
{Anstey}, D., {de Lera Acedo}, E., \& {Handley}, W. 2023, \mnras, 520, 850, \dodoi{10.1093/mnras/stad156}

\bibitem[{{Ashton} {et~al.}(2022){Ashton}, {Bernstein}, {Buchner}, {Chen}, {Cs{\'a}nyi}, {Fowlie}, {Feroz}, {Griffiths}, {Handley}, {Habeck}, {Higson}, {Hobson}, {Lasenby}, {Parkinson}, {P{\'a}rtay}, {Pitkin}, {Schneider}, {Speagle}, {South}, {Veitch}, {Wacker}, {Wales}, \& {Yallup}}]{Ashton22}
{Ashton}, G., {Bernstein}, N., {Buchner}, J., {et~al.} 2022, Nature Reviews Methods Primers, 2, 39, \dodoi{10.1038/s43586-022-00121-x}

\bibitem[{{Bale} {et~al.}(2023){Bale}, {Bassett}, {Burns}, {Dorigo Jones}, {Goetz}, {Hellum-Bye}, {Hermann}, {Hibbard}, {Maksimovic}, {McLean}, {Monsalve}, {O'Connor}, {Parsons}, {Pulupa}, {Pund}, {Rapetti}, {Rotermund}, {Saliwanchik}, {Slosar}, {Sundkvist}, \& {Suzuki}}]{LuSEENight}
{Bale}, S.~D., {Bassett}, N., {Burns}, J.~O., {et~al.} 2023, arXiv e-prints, arXiv:2301.10345, \dodoi{10.48550/arXiv.2301.10345}

\bibitem[{{Bassett} {et~al.}(2020){Bassett}, {Rapetti}, {Burns}, {Tauscher}, \& {MacDowall}}]{Bassett20}
{Bassett}, N., {Rapetti}, D., {Burns}, J.~O., {Tauscher}, K., \& {MacDowall}, R. 2020, Advances in Space Research, 66, 1265, \dodoi{10.1016/j.asr.2020.05.050}

\bibitem[{{Bassett} {et~al.}(2021){Bassett}, {Rapetti}, {Tauscher}, {Nhan}, {Bordenave}, {Hibbard}, \& {Burns}}]{Bassett21}
{Bassett}, N., {Rapetti}, D., {Tauscher}, K., {et~al.} 2021, \apj, 923, 33, \dodoi{10.3847/1538-4357/ac1cde}

\bibitem[{{Bera} {et~al.}(2023){Bera}, {Ghara}, {Chatterjee}, {Datta}, \& {Samui}}]{Bera23}
{Bera}, A., {Ghara}, R., {Chatterjee}, A., {Datta}, K.~K., \& {Samui}, S. 2023, Journal of Astrophysics and Astronomy, 44, 10, \dodoi{10.1007/s12036-022-09904-w}

\bibitem[{{Bernardi} {et~al.}(2016){Bernardi}, {Zwart}, {Price}, {Greenhill}, {Mesinger}, {Dowell}, {Eftekhari}, {Ellingson}, {Kocz}, \& {Schinzel}}]{Bernardi16}
{Bernardi}, G., {Zwart}, J.~T.~L., {Price}, D., {et~al.} 2016, \mnras, 461, 2847, \dodoi{10.1093/mnras/stw1499}

\bibitem[{{Bevins} {et~al.}(2022{\natexlab{a}}){Bevins}, {de Lera Acedo}, {Fialkov}, {Handley}, {Singh}, {Subrahmanyan}, \& {Barkana}}]{Bevins22a}
{Bevins}, H.~T.~J., {de Lera Acedo}, E., {Fialkov}, A., {et~al.} 2022{\natexlab{a}}, \mnras, 513, 4507, \dodoi{10.1093/mnras/stac1158}

\bibitem[{{Bevins} {et~al.}(2022{\natexlab{b}}){Bevins}, {Fialkov}, {de Lera Acedo}, {Handley}, {Singh}, {Subrahmanyan}, \& {Barkana}}]{Bevins22b}
{Bevins}, H.~T.~J., {Fialkov}, A., {de Lera Acedo}, E., {et~al.} 2022{\natexlab{b}}, Nature Astronomy, 6, 1473, \dodoi{10.1038/s41550-022-01825-6}

\bibitem[{{Bevins} {et~al.}(2021){Bevins}, {Handley}, {Fialkov}, {de Lera Acedo}, \& {Javid}}]{globalemu}
{Bevins}, H.~T.~J., {Handley}, W.~J., {Fialkov}, A., {de Lera Acedo}, E., \& {Javid}, K. 2021, \mnras, 508, 2923, \dodoi{10.1093/mnras/stab2737}

\bibitem[{{Bevins} {et~al.}(2023){Bevins}, {Heimersheim}, {Abril-Cabezas}, {Fialkov}, {de Lera Acedo}, {Handley}, {Singh}, \& {Barkana}}]{Bevins23}
{Bevins}, H. T.~J., {Heimersheim}, S., {Abril-Cabezas}, I., {et~al.} 2023, arXiv e-prints, arXiv:2301.03298, \dodoi{10.48550/arXiv.2301.03298}

\bibitem[{{Bond} {et~al.}(1991){Bond}, {Cole}, {Efstathiou}, \& {Kaiser}}]{Bond91}
{Bond}, J.~R., {Cole}, S., {Efstathiou}, G., \& {Kaiser}, N. 1991, \apj, 379, 440, \dodoi{10.1086/170520}

\bibitem[{{Bouwens} {et~al.}(2023){Bouwens}, {Illingworth}, {Oesch}, {Stefanon}, {Naidu}, {van Leeuwen}, \& {Magee}}]{Bouwens23}
{Bouwens}, R., {Illingworth}, G., {Oesch}, P., {et~al.} 2023, \mnras, 523, 1009, \dodoi{10.1093/mnras/stad1014}

\bibitem[{{Bouwens} {et~al.}(2015){Bouwens}, {Illingworth}, {Oesch}, {Trenti}, {Labb{\'e}}, {Bradley}, {Carollo}, {van Dokkum}, {Gonzalez}, {Holwerda}, {Franx}, {Spitler}, {Smit}, \& {Magee}}]{Bouwens15}
{Bouwens}, R.~J., {Illingworth}, G.~D., {Oesch}, P.~A., {et~al.} 2015, \apj, 803, 34, \dodoi{10.1088/0004-637X/803/1/34}

\bibitem[{{Bowman} {et~al.}(2018){Bowman}, {Rogers}, {Monsalve}, {Mozdzen}, \& {Mahesh}}]{EDGES}
{Bowman}, J.~D., {Rogers}, A. E.~E., {Monsalve}, R.~A., {Mozdzen}, T.~J., \& {Mahesh}, N. 2018, \nat, 555, 67, \dodoi{10.1038/nature25792}

\bibitem[{{Boylan-Kolchin}(2023)}]{Boylan-Kolchin23}
{Boylan-Kolchin}, M. 2023, Nature Astronomy, 7, 731, \dodoi{10.1038/s41550-023-01937-7}

\bibitem[{{Bradley} {et~al.}(2019){Bradley}, {Tauscher}, {Rapetti}, \& {Burns}}]{Bradley19}
{Bradley}, R.~F., {Tauscher}, K., {Rapetti}, D., \& {Burns}, J.~O. 2019, \apj, 874, 153, \dodoi{10.3847/1538-4357/ab0d8b}

\bibitem[{{Breitman} {et~al.}(2023){Breitman}, {Mesinger}, {Murray}, {Prelogovic}, {Qin}, \& {Trotta}}]{Breitman23}
{Breitman}, D., {Mesinger}, A., {Murray}, S., {et~al.} 2023, arXiv e-prints, arXiv:2309.05697, \dodoi{10.48550/arXiv.2309.05697}

\bibitem[{{Buchner}(2016)}]{Buchner16}
{Buchner}, J. 2016, Statistics and Computing, 26, 383, \dodoi{10.1007/s11222-014-9512-y}

\bibitem[{{Buchner}(2019)}]{Buchner19}
---. 2019, \pasp, 131, 108005, \dodoi{10.1088/1538-3873/aae7fc}

\bibitem[{{Buchner}(2021)}]{UltraNest}
---. 2021, The Journal of Open Source Software, 6, 3001, \dodoi{10.21105/joss.03001}

\bibitem[{{Buchner}(2023)}]{Buchner23}
---. 2023, Statistics Surveys, 17, 169, \dodoi{10.1214/23-SS144}

\bibitem[{{Burns} {et~al.}(2021{\natexlab{a}}){Burns}, {Hallinan}, {Chang}, {Anderson}, {Bowman}, {Bradley}, {Furlanetto}, {Hegedus}, {Kasper}, {Kocz}, {Lazio}, {Lux}, {MacDowall}, {Mirocha}, {Nesnas}, {Pober}, {Polidan}, {Rapetti}, {Romero-Wolf}, {Slosar}, {Stebbins}, {Teitelbaum}, \& {White}}]{FARSIDEBurns21}
{Burns}, J., {Hallinan}, G., {Chang}, T.-C., {et~al.} 2021{\natexlab{a}}, arXiv e-prints, arXiv:2103.08623, \dodoi{10.48550/arXiv.2103.08623}

\bibitem[{{Burns} {et~al.}(2021{\natexlab{b}}){Burns}, {MacDowall}, {Bale}, {Hallinan}, {Bassett}, \& {Hegedus}}]{Burns21}
{Burns}, J.~O., {MacDowall}, R., {Bale}, S., {et~al.} 2021{\natexlab{b}}, \psj, 2, 44, \dodoi{10.3847/PSJ/abdfc3}

\bibitem[{{Bye} {et~al.}(2022){Bye}, {Portillo}, \& {Fialkov}}]{21cmVAE}
{Bye}, C.~H., {Portillo}, S. K.~N., \& {Fialkov}, A. 2022, \apj, 930, 79, \dodoi{10.3847/1538-4357/ac6424}

\bibitem[{{Chatterjee} {et~al.}(2021){Chatterjee}, {Choudhury}, \& {Mitra}}]{Chatterjee21}
{Chatterjee}, A., {Choudhury}, T.~R., \& {Mitra}, S. 2021, \mnras, 507, 2405, \dodoi{10.1093/mnras/stab2316}

\bibitem[{{Cohen} {et~al.}(2020){Cohen}, {Fialkov}, {Barkana}, \& {Monsalve}}]{Cohen20}
{Cohen}, A., {Fialkov}, A., {Barkana}, R., \& {Monsalve}, R.~A. 2020, \mnras, 495, 4845, \dodoi{10.1093/mnras/staa1530}

\bibitem[{{Cranmer} {et~al.}(2020){Cranmer}, {Brehmer}, \& {Louppe}}]{Cranmer20}
{Cranmer}, K., {Brehmer}, J., \& {Louppe}, G. 2020, Proceedings of the National Academy of Science, 117, 30055, \dodoi{10.1073/pnas.1912789117}

\bibitem[{{Das} {et~al.}(2017){Das}, {Mesinger}, {Pallottini}, {Ferrara}, \& {Wise}}]{Das17}
{Das}, A., {Mesinger}, A., {Pallottini}, A., {Ferrara}, A., \& {Wise}, J.~H. 2017, \mnras, 469, 1166, \dodoi{10.1093/mnras/stx943}

\bibitem[{{de Lera Acedo} {et~al.}(2022){de Lera Acedo}, {de Villiers}, {Razavi-Ghods}, {Handley}, {Fialkov}, {Magro}, {Anstey}, {Bevins}, {Chiello}, {Cumner}, {Josaitis}, {Roque}, {Sims}, {Scheutwinkel}, {Alexander}, {Bernardi}, {Carey}, {Cavillot}, {Croukamp}, {Ely}, {Gessey-Jones}, {Gueuning}, {Hills}, {Kulkarni}, {Maiolino}, {Meerburg}, {Mittal}, {Pritchard}, {Puchwein}, {Saxena}, {Shen}, {Smirnov}, {Spinelli}, \& {Zarb-Adami}}]{REACH}
{de Lera Acedo}, E., {de Villiers}, D.~I.~L., {Razavi-Ghods}, N., {et~al.} 2022, Nature Astronomy, 6, 984, \dodoi{10.1038/s41550-022-01709-9}

\bibitem[{{Donnan} {et~al.}(2023){Donnan}, {McLeod}, {Dunlop}, {McLure}, {Carnall}, {Begley}, {Cullen}, {Hamadouche}, {Bowler}, {Magee}, {McCracken}, {Milvang-Jensen}, {Moneti}, \& {Targett}}]{Donnan23}
{Donnan}, C.~T., {McLeod}, D.~J., {Dunlop}, J.~S., {et~al.} 2023, \mnras, 518, 6011, \dodoi{10.1093/mnras/stac3472}

\bibitem[{{Eldridge} \& {Stanway}(2009)}]{Eldridge09}
{Eldridge}, J.~J., \& {Stanway}, E.~R. 2009, \mnras, 400, 1019, \dodoi{10.1111/j.1365-2966.2009.15514.x}

\bibitem[{{Feroz} \& {Hobson}(2008)}]{Feroz08}
{Feroz}, F., \& {Hobson}, M.~P. 2008, \mnras, 384, 449, \dodoi{10.1111/j.1365-2966.2007.12353.x}

\bibitem[{{Feroz} {et~al.}(2009){Feroz}, {Hobson}, \& {Bridges}}]{Feroz09}
{Feroz}, F., {Hobson}, M.~P., \& {Bridges}, M. 2009, \mnras, 398, 1601, \dodoi{10.1111/j.1365-2966.2009.14548.x}

\bibitem[{{Feroz} {et~al.}(2019){Feroz}, {Hobson}, {Cameron}, \& {Pettitt}}]{Feroz19}
{Feroz}, F., {Hobson}, M.~P., {Cameron}, E., \& {Pettitt}, A.~N. 2019, The Open Journal of Astrophysics, 2, 10, \dodoi{10.21105/astro.1306.2144}

\bibitem[{{Fialkov} \& {Barkana}(2014)}]{FialkovBarkana14}
{Fialkov}, A., \& {Barkana}, R. 2014, \mnras, 445, 213, \dodoi{10.1093/mnras/stu1744}

\bibitem[{{Finkelstein} {et~al.}(2023){Finkelstein}, {Bagley}, {Ferguson}, {Wilkins}, {Kartaltepe}, {Papovich}, {Yung}, {Haro}, {Behroozi}, {Dickinson}, {Kocevski}, {Koekemoer}, {Larson}, {Le Bail}, {Morales}, {P{\'e}rez-Gonz{\'a}lez}, {Burgarella}, {Dav{\'e}}, {Hirschmann}, {Somerville}, {Wuyts}, {Bromm}, {Casey}, {Fontana}, {Fujimoto}, {Gardner}, {Giavalisco}, {Grazian}, {Grogin}, {Hathi}, {Hutchison}, {Jha}, {Jogee}, {Kewley}, {Kirkpatrick}, {Long}, {Lotz}, {Pentericci}, {Pierel}, {Pirzkal}, {Ravindranath}, {Ryan}, {Trump}, {Yang}, {Bhatawdekar}, {Bisigello}, {Buat}, {Calabr{\`o}}, {Castellano}, {Cleri}, {Cooper}, {Croton}, {Daddi}, {Dekel}, {Elbaz}, {Franco}, {Gawiser}, {Holwerda}, {Huertas-Company}, {Jaskot}, {Leung}, {Lucas}, {Mobasher}, {Pandya}, {Tacchella}, {Weiner}, \& {Zavala}}]{Finkelstein23}
{Finkelstein}, S.~L., {Bagley}, M.~B., {Ferguson}, H.~C., {et~al.} 2023, \apjl, 946, L13, \dodoi{10.3847/2041-8213/acade4}

\bibitem[{{Foreman-Mackey}(2016)}]{corner}
{Foreman-Mackey}, D. 2016, The Journal of Open Source Software, 1, 24, \dodoi{10.21105/joss.00024}

\bibitem[{{Foreman-Mackey} {et~al.}(2013){Foreman-Mackey}, {Hogg}, {Lang}, \& {Goodman}}]{emcee}
{Foreman-Mackey}, D., {Hogg}, D.~W., {Lang}, D., \& {Goodman}, J. 2013, \pasp, 125, 306, \dodoi{10.1086/670067}

\bibitem[{{Furlanetto} {et~al.}(2017){Furlanetto}, {Mirocha}, {Mebane}, \& {Sun}}]{Furlanetto17}
{Furlanetto}, S.~R., {Mirocha}, J., {Mebane}, R.~H., \& {Sun}, G. 2017, \mnras, 472, 1576, \dodoi{10.1093/mnras/stx2132}

\bibitem[{{Furlanetto} {et~al.}(2006){Furlanetto}, {Oh}, \& {Briggs}}]{Furlanetto06}
{Furlanetto}, S.~R., {Oh}, S.~P., \& {Briggs}, F.~H. 2006, \physrep, 433, 181, \dodoi{10.1016/j.physrep.2006.08.002}

\bibitem[{{Garsden} {et~al.}(2021){Garsden}, {Greenhill}, {Bernardi}, {Fialkov}, {Price}, {Mitchell}, {Dowell}, {Spinelli}, \& {Schinzel}}]{LEDA}
{Garsden}, H., {Greenhill}, L., {Bernardi}, G., {et~al.} 2021, \mnras, 506, 5802, \dodoi{10.1093/mnras/stab1671}

\bibitem[{{Ghara} {et~al.}(2015){Ghara}, {Choudhury}, \& {Datta}}]{Ghara15}
{Ghara}, R., {Choudhury}, T.~R., \& {Datta}, K.~K. 2015, \mnras, 447, 1806, \dodoi{10.1093/mnras/stu2512}

\bibitem[{{Ghara} {et~al.}(2018){Ghara}, {Mellema}, {Giri}, {Choudhury}, {Datta}, \& {Majumdar}}]{Ghara18}
{Ghara}, R., {Mellema}, G., {Giri}, S.~K., {et~al.} 2018, \mnras, 476, 1741, \dodoi{10.1093/mnras/sty314}

\bibitem[{{Handley} {et~al.}(2015{\natexlab{a}}){Handley}, {Hobson}, \& {Lasenby}}]{Handley15a}
{Handley}, W.~J., {Hobson}, M.~P., \& {Lasenby}, A.~N. 2015{\natexlab{a}}, \mnras, 450, L61, \dodoi{10.1093/mnrasl/slv047}

\bibitem[{{Handley} {et~al.}(2015{\natexlab{b}}){Handley}, {Hobson}, \& {Lasenby}}]{Handley15b}
---. 2015{\natexlab{b}}, \mnras, 453, 4384, \dodoi{10.1093/mnras/stv1911}

\bibitem[{{Harikane} {et~al.}(2023){Harikane}, {Ouchi}, {Oguri}, {Ono}, {Nakajima}, {Isobe}, {Umeda}, {Mawatari}, \& {Zhang}}]{Harikane23}
{Harikane}, Y., {Ouchi}, M., {Oguri}, M., {et~al.} 2023, \apjs, 265, 5, \dodoi{10.3847/1538-4365/acaaa9}

\bibitem[{{Harris} {et~al.}(2020){Harris}, {Millman}, {van der Walt}, {Gommers}, {Virtanen}, {Cournapeau}, {Wieser}, {Taylor}, {Berg}, {Smith}, {Kern}, {Picus}, {Hoyer}, {van Kerkwijk}, {Brett}, {Haldane}, {del R{\'\i}o}, {Wiebe}, {Peterson}, {G{\'e}rard-Marchant}, {Sheppard}, {Reddy}, {Weckesser}, {Abbasi}, {Gohlke}, \& {Oliphant}}]{numpy}
{Harris}, C.~R., {Millman}, K.~J., {van der Walt}, S.~J., {et~al.} 2020, \nat, 585, 357, \dodoi{10.1038/s41586-020-2649-2}

\bibitem[{{Heavens}(2016)}]{Heavens16}
{Heavens}, A. 2016, Entropy, 18, 236, \dodoi{10.3390/e18060236}

\bibitem[{{Hibbard} {et~al.}(2022){Hibbard}, {Mirocha}, {Rapetti}, {Bassett}, {Burns}, \& {Tauscher}}]{Hibbard22}
{Hibbard}, J.~J., {Mirocha}, J., {Rapetti}, D., {et~al.} 2022, \apj, 929, 151, \dodoi{10.3847/1538-4357/ac5ea3}

\bibitem[{{Hibbard} {et~al.}(2023){Hibbard}, {Rapetti}, {Burns}, {Mahesh}, \& {Bassett}}]{Hibbard23}
{Hibbard}, J.~J., {Rapetti}, D., {Burns}, J.~O., {Mahesh}, N., \& {Bassett}, N. 2023, arXiv e-prints, arXiv:2304.09959, \dodoi{10.48550/arXiv.2304.09959}

\bibitem[{{Hibbard} {et~al.}(2020){Hibbard}, {Tauscher}, {Rapetti}, \& {Burns}}]{Hibbard20}
{Hibbard}, J.~J., {Tauscher}, K., {Rapetti}, D., \& {Burns}, J.~O. 2020, \apj, 905, 113, \dodoi{10.3847/1538-4357/abc3c5}

\bibitem[{{Hills} {et~al.}(2018){Hills}, {Kulkarni}, {Meerburg}, \& {Puchwein}}]{Hills18}
{Hills}, R., {Kulkarni}, G., {Meerburg}, P.~D., \& {Puchwein}, E. 2018, \nat, 564, E32, \dodoi{10.1038/s41586-018-0796-5}

\bibitem[{{Hobson} {et~al.}(2002){Hobson}, {Bridle}, \& {Lahav}}]{Hobson02}
{Hobson}, M.~P., {Bridle}, S.~L., \& {Lahav}, O. 2002, \mnras, 335, 377, \dodoi{10.1046/j.1365-8711.2002.05614.x}

\bibitem[{{Hunter}(2007)}]{matplotlib}
{Hunter}, J.~D. 2007, Computing in Science and Engineering, 9, 90, \dodoi{10.1109/MCSE.2007.55}

\bibitem[{{Hutter} {et~al.}(2023){Hutter}, {Trebitsch}, {Dayal}, {Gottl{\"o}ber}, {Yepes}, \& {Legrand}}]{Hutter23}
{Hutter}, A., {Trebitsch}, M., {Dayal}, P., {et~al.} 2023, \mnras, 524, 6124, \dodoi{10.1093/mnras/stad2230}

\bibitem[{Jeffreys(1998)}]{jeffreys1998}
Jeffreys, H. 1998, The Theory of Probability, Oxford Classic Texts in the Physical Sciences (OUP Oxford).
\newblock \url{https://books.google.com/books?id=vh9Act9rtzQC}

\bibitem[{Kass \& Raftery(1995)}]{KassRaftery95}
Kass, R.~E., \& Raftery, A.~E. 1995, Journal of the American Statistical Association, 90, 773, \dodoi{10.1080/01621459.1995.10476572}

\bibitem[{{Kern} {et~al.}(2017){Kern}, {Liu}, {Parsons}, {Mesinger}, \& {Greig}}]{Kern17}
{Kern}, N.~S., {Liu}, A., {Parsons}, A.~R., {Mesinger}, A., \& {Greig}, B. 2017, \apj, 848, 23, \dodoi{10.3847/1538-4357/aa8bb4}

\bibitem[{{Kern} {et~al.}(2020){Kern}, {Parsons}, {Dillon}, {Lanman}, {Liu}, {Bull}, {Ewall-Wice}, {Abdurashidova}, {Aguirre}, {Alexander}, {Ali}, {Balfour}, {Beardsley}, {Bernardi}, {Bowman}, {Bradley}, {Burba}, {Carilli}, {Cheng}, {DeBoer}, {Dexter}, {de Lera Acedo}, {Fagnoni}, {Fritz}, {Furlanetto}, {Glendenning}, {Gorthi}, {Greig}, {Grobbelaar}, {Halday}, {Hazelton}, {Hewitt}, {Hickish}, {Jacobs}, {Julius}, {Kerrigan}, {Kittiwisit}, {Kohn}, {Kolopanis}, {La Plante}, {Lekalake}, {MacMahon}, {Malan}, {Malgas}, {Maree}, {Martinot}, {Matsetela}, {Mesinger}, {Molewa}, {Morales}, {Mosiane}, {Murray}, {Neben}, {Parsons}, {Patra}, {Pieterse}, {Pober}, {Razavi-Ghods}, {Ringuette}, {Robnett}, {Rosie}, {Sims}, {Smith}, {Syce}, {Thyagarajan}, {Williams}, \& {Zheng}}]{Kern20}
{Kern}, N.~S., {Parsons}, A.~R., {Dillon}, J.~S., {et~al.} 2020, \apj, 888, 70, \dodoi{10.3847/1538-4357/ab5e8a}

\bibitem[{Kluyver {et~al.}(2016)Kluyver, Ragan-Kelley, P{\'e}rez, Granger, Bussonnier, Frederic, Kelley, Hamrick, Grout, Corlay, Ivanov, Avila, Abdalla, Willing, \& development team}]{jupyter}
Kluyver, T., Ragan-Kelley, B., P{\'e}rez, F., {et~al.} 2016, in Positioning and Power in Academic Publishing: Players, Agents and Agendas, ed. F.~Loizides \& B.~Scmidt (Netherlands: IOS Press), 87--90.
\newblock \url{https://eprints.soton.ac.uk/403913/}

\bibitem[{{Lahav} {et~al.}(2000){Lahav}, {Bridle}, {Hobson}, {Lasenby}, \& {Sodr{\'e}}}]{Lahav00}
{Lahav}, O., {Bridle}, S.~L., {Hobson}, M.~P., {Lasenby}, A.~N., \& {Sodr{\'e}}, L. 2000, \mnras, 315, L45, \dodoi{10.1046/j.1365-8711.2000.03633.x}

\bibitem[{{Leeney} {et~al.}(2022){Leeney}, {Handley}, \& {de Lera Acedo}}]{Leeney22}
{Leeney}, S.~A.~K., {Handley}, W.~J., \& {de Lera Acedo}, E. 2022, arXiv e-prints, arXiv:2211.15448, \dodoi{10.48550/arXiv.2211.15448}

\bibitem[{{Lemos} {et~al.}(2023){Lemos}, {Weaverdyck}, {Rollins}, {Muir}, {Fert{\'e}}, {Liddle}, {Campos}, {Huterer}, {Raveri}, {Zuntz}, {Di Valentino}, {Fang}, {Hartley}, {Aguena}, {Allam}, {Annis}, {Bertin}, {Bocquet}, {Brooks}, {Burke}, {Carnero Rosell}, {Carrasco Kind}, {Carretero}, {Castander}, {Choi}, {Costanzi}, {Crocce}, {da Costa}, {Pereira}, {Dietrich}, {Everett}, {Ferrero}, {Frieman}, {Garc{\'\i}a-Bellido}, {Gatti}, {Gaztanaga}, {Gerdes}, {Gruen}, {Gruendl}, {Gschwend}, {Gutierrez}, {Hinton}, {Hollowood}, {Honscheid}, {James}, {Kuehn}, {Kuropatkin}, {Lima}, {March}, {Melchior}, {Menanteau}, {Miquel}, {Morgan}, {Palmese}, {Paz-Chinch{\'o}n}, {Pieres}, {Malag{\'o}n}, {Porredon}, {Sanchez}, {Scarpine}, {Schubnell}, {Serrano}, {Sevilla-Noarbe}, {Smith}, {Suchyta}, {Swanson}, {Tarle}, {Thomas}, {To}, {Varga}, {Weller}, \& {DES Collaboration}}]{Lemos23}
{Lemos}, P., {Weaverdyck}, N., {Rollins}, R.~P., {et~al.} 2023, \mnras, 521, 1184, \dodoi{10.1093/mnras/stac2786}

\bibitem[{{Liu} {et~al.}(2013){Liu}, {Pritchard}, {Tegmark}, \& {Loeb}}]{Liu13}
{Liu}, A., {Pritchard}, J.~R., {Tegmark}, M., \& {Loeb}, A. 2013, \prd, 87, 043002, \dodoi{10.1103/PhysRevD.87.043002}

\bibitem[{{Liu} \& {Shaw}(2020)}]{Liu20}
{Liu}, A., \& {Shaw}, J.~R. 2020, \pasp, 132, 062001, \dodoi{10.1088/1538-3873/ab5bfd}

\bibitem[{{Lovell} {et~al.}(2023){Lovell}, {Harrison}, {Harikane}, {Tacchella}, \& {Wilkins}}]{Lovell23}
{Lovell}, C.~C., {Harrison}, I., {Harikane}, Y., {Tacchella}, S., \& {Wilkins}, S.~M. 2023, \mnras, 518, 2511, \dodoi{10.1093/mnras/stac3224}

\bibitem[{{Madau} {et~al.}(1997){Madau}, {Meiksin}, \& {Rees}}]{Madau97}
{Madau}, P., {Meiksin}, A., \& {Rees}, M.~J. 1997, \apj, 475, 429, \dodoi{10.1086/303549}

\bibitem[{{Mason} {et~al.}(2023{\natexlab{a}}){Mason}, {Mu{\~n}oz}, {Greig}, {Mesinger}, \& {Park}}]{Mason23Fish}
{Mason}, C.~A., {Mu{\~n}oz}, J.~B., {Greig}, B., {Mesinger}, A., \& {Park}, J. 2023{\natexlab{a}}, \mnras, \dodoi{10.1093/mnras/stad2145}

\bibitem[{{Mason} {et~al.}(2023{\natexlab{b}}){Mason}, {Trenti}, \& {Treu}}]{Mason23}
{Mason}, C.~A., {Trenti}, M., \& {Treu}, T. 2023{\natexlab{b}}, \mnras, 521, 497, \dodoi{10.1093/mnras/stad035}

\bibitem[{{Mertens} {et~al.}(2020){Mertens}, {Mevius}, {Koopmans}, {Offringa}, {Mellema}, {Zaroubi}, {Brentjens}, {Gan}, {Gehlot}, {Pandey}, {Sardarabadi}, {Vedantham}, {Yatawatta}, {Asad}, {Ciardi}, {Chapman}, {Gazagnes}, {Ghara}, {Ghosh}, {Giri}, {Iliev}, {Jeli{\'c}}, {Kooistra}, {Mondal}, {Schaye}, \& {Silva}}]{LOFAR}
{Mertens}, F.~G., {Mevius}, M., {Koopmans}, L.~V.~E., {et~al.} 2020, \mnras, 493, 1662, \dodoi{10.1093/mnras/staa327}

\bibitem[{{Mesinger} {et~al.}(2011){Mesinger}, {Furlanetto}, \& {Cen}}]{Mesinger11}
{Mesinger}, A., {Furlanetto}, S., \& {Cen}, R. 2011, \mnras, 411, 955, \dodoi{10.1111/j.1365-2966.2010.17731.x}

\bibitem[{{Mineo} {et~al.}(2012){Mineo}, {Gilfanov}, \& {Sunyaev}}]{Mineo12}
{Mineo}, S., {Gilfanov}, M., \& {Sunyaev}, R. 2012, \mnras, 419, 2095, \dodoi{10.1111/j.1365-2966.2011.19862.x}

\bibitem[{{Mirocha}(2014)}]{Mirocha14}
{Mirocha}, J. 2014, \mnras, 443, 1211, \dodoi{10.1093/mnras/stu1193}

\bibitem[{{Mirocha} \& {Furlanetto}(2019)}]{Mirocha19}
{Mirocha}, J., \& {Furlanetto}, S.~R. 2019, \mnras, 483, 1980, \dodoi{10.1093/mnras/sty3260}

\bibitem[{{Mirocha} \& {Furlanetto}(2023)}]{Mirocha23}
---. 2023, \mnras, 519, 843, \dodoi{10.1093/mnras/stac3578}

\bibitem[{{Mirocha} {et~al.}(2017){Mirocha}, {Furlanetto}, \& {Sun}}]{Mirocha17}
{Mirocha}, J., {Furlanetto}, S.~R., \& {Sun}, G. 2017, \mnras, 464, 1365, \dodoi{10.1093/mnras/stw2412}

\bibitem[{{Mirocha} {et~al.}(2021){Mirocha}, {La Plante}, \& {Liu}}]{Mirocha21}
{Mirocha}, J., {La Plante}, P., \& {Liu}, A. 2021, \mnras, 507, 3872, \dodoi{10.1093/mnras/stab1871}

\bibitem[{{Mirocha} {et~al.}(2012){Mirocha}, {Skory}, {Burns}, \& {Wise}}]{Mirocha12}
{Mirocha}, J., {Skory}, S., {Burns}, J.~O., \& {Wise}, J.~H. 2012, \apj, 756, 94, \dodoi{10.1088/0004-637X/756/1/94}

\bibitem[{{Mitsuda} {et~al.}(1984){Mitsuda}, {Inoue}, {Koyama}, {Makishima}, {Matsuoka}, {Ogawara}, {Shibazaki}, {Suzuki}, {Tanaka}, \& {Hirano}}]{Mitsuda84}
{Mitsuda}, K., {Inoue}, H., {Koyama}, K., {et~al.} 1984, \pasj, 36, 741

\bibitem[{{Monsalve} {et~al.}(2019){Monsalve}, {Fialkov}, {Bowman}, {Rogers}, {Mozdzen}, {Cohen}, {Barkana}, \& {Mahesh}}]{Monsalve19}
{Monsalve}, R.~A., {Fialkov}, A., {Bowman}, J.~D., {et~al.} 2019, \apj, 875, 67, \dodoi{10.3847/1538-4357/ab07be}

\bibitem[{{Monsalve} {et~al.}(2018){Monsalve}, {Greig}, {Bowman}, {Mesinger}, {Rogers}, {Mozdzen}, {Kern}, \& {Mahesh}}]{Monsalve18}
{Monsalve}, R.~A., {Greig}, B., {Bowman}, J.~D., {et~al.} 2018, \apj, 863, 11, \dodoi{10.3847/1538-4357/aace54}

\bibitem[{{Mu{\~n}oz} {et~al.}(2020){Mu{\~n}oz}, {Dvorkin}, \& {Cyr-Racine}}]{Munoz20}
{Mu{\~n}oz}, J.~B., {Dvorkin}, C., \& {Cyr-Racine}, F.-Y. 2020, \prd, 101, 063526, \dodoi{10.1103/PhysRevD.101.063526}

\bibitem[{{Murray} {et~al.}(2020){Murray}, {Greig}, {Mesinger}, {Mu{\~n}oz}, {Qin}, {Park}, \& {Watkinson}}]{Murray20}
{Murray}, S., {Greig}, B., {Mesinger}, A., {et~al.} 2020, The Journal of Open Source Software, 5, 2582, \dodoi{10.21105/joss.02582}

\bibitem[{{Murray} {et~al.}(2022){Murray}, {Bowman}, {Sims}, {Mahesh}, {Rogers}, {Monsalve}, {Samson}, \& {Vydula}}]{Murray22}
{Murray}, S.~G., {Bowman}, J.~D., {Sims}, P.~H., {et~al.} 2022, \mnras, 517, 2264, \dodoi{10.1093/mnras/stac2600}

\bibitem[{{Murray} {et~al.}(2013){Murray}, {Power}, \& {Robotham}}]{Murray13}
{Murray}, S.~G., {Power}, C., \& {Robotham}, A.~S.~G. 2013, Astronomy and Computing, 3, 23, \dodoi{10.1016/j.ascom.2013.11.001}

\bibitem[{{Naidu} {et~al.}(2022){Naidu}, {Oesch}, {van Dokkum}, {Nelson}, {Suess}, {Brammer}, {Whitaker}, {Illingworth}, {Bouwens}, {Tacchella}, {Matthee}, {Allen}, {Bezanson}, {Conroy}, {Labbe}, {Leja}, {Leonova}, {Magee}, {Price}, {Setton}, {Strait}, {Stefanon}, {Toft}, {Weaver}, \& {Weibel}}]{Naidu22}
{Naidu}, R.~P., {Oesch}, P.~A., {van Dokkum}, P., {et~al.} 2022, \apjl, 940, L14, \dodoi{10.3847/2041-8213/ac9b22}

\bibitem[{{Paciga} {et~al.}(2011){Paciga}, {Chang}, {Gupta}, {Nityanada}, {Odegova}, {Pen}, {Peterson}, {Roy}, \& {Sigurdson}}]{GMRT}
{Paciga}, G., {Chang}, T.-C., {Gupta}, Y., {et~al.} 2011, \mnras, 413, 1174, \dodoi{10.1111/j.1365-2966.2011.18208.x}

\bibitem[{{Pagano} {et~al.}(2022){Pagano}, {Sims}, {Liu}, {Anstey}, {Handley}, \& {De Lera Acedo}}]{Pagano22}
{Pagano}, M., {Sims}, P., {Liu}, A., {et~al.} 2022, arXiv e-prints, arXiv:2211.10448, \dodoi{10.48550/arXiv.2211.10448}

\bibitem[{{Planck Collaboration} {et~al.}(2020){Planck Collaboration}, {Aghanim}, {Akrami}, {Ashdown}, {Aumont}, {Baccigalupi}, {Ballardini}, {Banday}, {Barreiro}, {Bartolo}, {Basak}, {Battye}, {Benabed}, {Bernard}, {Bersanelli}, {Bielewicz}, {Bock}, {Bond}, {Borrill}, {Bouchet}, {Boulanger}, {Bucher}, {Burigana}, {Butler}, {Calabrese}, {Cardoso}, {Carron}, {Challinor}, {Chiang}, {Chluba}, {Colombo}, {Combet}, {Contreras}, {Crill}, {Cuttaia}, {de Bernardis}, {de Zotti}, {Delabrouille}, {Delouis}, {Di Valentino}, {Diego}, {Dor{\'e}}, {Douspis}, {Ducout}, {Dupac}, {Dusini}, {Efstathiou}, {Elsner}, {En{\ss}lin}, {Eriksen}, {Fantaye}, {Farhang}, {Fergusson}, {Fernandez-Cobos}, {Finelli}, {Forastieri}, {Frailis}, {Fraisse}, {Franceschi}, {Frolov}, {Galeotta}, {Galli}, {Ganga}, {G{\'e}nova-Santos}, {Gerbino}, {Ghosh}, {Gonz{\'a}lez-Nuevo}, {G{\'o}rski}, {Gratton}, {Gruppuso}, {Gudmundsson}, {Hamann}, {Handley}, {Hansen}, {Herranz}, {Hildebrandt}, {Hivon}, {Huang}, {Jaffe}, {Jones}, {Karakci}, {Keih{\"a}nen},
  {Keskitalo}, {Kiiveri}, {Kim}, {Kisner}, {Knox}, {Krachmalnicoff}, {Kunz}, {Kurki-Suonio}, {Lagache}, {Lamarre}, {Lasenby}, {Lattanzi}, {Lawrence}, {Le Jeune}, {Lemos}, {Lesgourgues}, {Levrier}, {Lewis}, {Liguori}, {Lilje}, {Lilley}, {Lindholm}, {L{\'o}pez-Caniego}, {Lubin}, {Ma}, {Mac{\'\i}as-P{\'e}rez}, {Maggio}, {Maino}, {Mandolesi}, {Mangilli}, {Marcos-Caballero}, {Maris}, {Martin}, {Martinelli}, {Mart{\'\i}nez-Gonz{\'a}lez}, {Matarrese}, {Mauri}, {McEwen}, {Meinhold}, {Melchiorri}, {Mennella}, {Migliaccio}, {Millea}, {Mitra}, {Miville-Desch{\^e}nes}, {Molinari}, {Montier}, {Morgante}, {Moss}, {Natoli}, {N{\o}rgaard-Nielsen}, {Pagano}, {Paoletti}, {Partridge}, {Patanchon}, {Peiris}, {Perrotta}, {Pettorino}, {Piacentini}, {Polastri}, {Polenta}, {Puget}, {Rachen}, {Reinecke}, {Remazeilles}, {Renzi}, {Rocha}, {Rosset}, {Roudier}, {Rubi{\~n}o-Mart{\'\i}n}, {Ruiz-Granados}, {Salvati}, {Sandri}, {Savelainen}, {Scott}, {Shellard}, {Sirignano}, {Sirri}, {Spencer}, {Sunyaev}, {Suur-Uski}, {Tauber}, {Tavagnacco},
  {Tenti}, {Toffolatti}, {Tomasi}, {Trombetti}, {Valenziano}, {Valiviita}, {Van Tent}, {Vibert}, {Vielva}, {Villa}, {Vittorio}, {Wandelt}, {Wehus}, {White}, {White}, {Zacchei}, \& {Zonca}}]{Planck18}
{Planck Collaboration}, {Aghanim}, N., {Akrami}, Y., {et~al.} 2020, \aap, 641, A6, \dodoi{10.1051/0004-6361/201833910}

\bibitem[{{Polidan} {et~al.}(2022){Polidan}, {Lopez}, {Burns}, {Ignatiev}, {Carol}, \& {Curreri}}]{FarViewPoster}
{Polidan}, R., {Lopez}, J., J., {Burns}, J.~O., {et~al.} 2022, in American Astronomical Society Meeting Abstracts, Vol.~54, American Astronomical Society Meeting Abstracts, 206.10

\bibitem[{{Prelogovi{\'c}} \& {Mesinger}(2023)}]{PrelogovicMesinger23}
{Prelogovi{\'c}}, D., \& {Mesinger}, A. 2023, \mnras, \dodoi{10.1093/mnras/stad2027}

\bibitem[{{Press} \& {Schechter}(1974)}]{PressSchechter}
{Press}, W.~H., \& {Schechter}, P. 1974, \apj, 187, 425, \dodoi{10.1086/152650}

\bibitem[{{Qin} {et~al.}(2020){Qin}, {Mesinger}, {Park}, {Greig}, \& {Mu{\~n}oz}}]{Qin20}
{Qin}, Y., {Mesinger}, A., {Park}, J., {Greig}, B., \& {Mu{\~n}oz}, J.~B. 2020, \mnras, 495, 123, \dodoi{10.1093/mnras/staa1131}

\bibitem[{{Rapetti} {et~al.}(2020){Rapetti}, {Tauscher}, {Mirocha}, \& {Burns}}]{PaperII}
{Rapetti}, D., {Tauscher}, K., {Mirocha}, J., \& {Burns}, J.~O. 2020, \apj, 897, 174, \dodoi{10.3847/1538-4357/ab9b29}

\bibitem[{{Ryan} {et~al.}(2023){Ryan}, {Stevenson}, {Trendafilova}, \& {Meyers}}]{Ryan23}
{Ryan}, J., {Stevenson}, B., {Trendafilova}, C., \& {Meyers}, J. 2023, \prd, 107, 103506, \dodoi{10.1103/PhysRevD.107.103506}

\bibitem[{{Santos} {et~al.}(2010){Santos}, {Ferramacho}, {Silva}, {Amblard}, \& {Cooray}}]{Santos10}
{Santos}, M.~G., {Ferramacho}, L., {Silva}, M.~B., {Amblard}, A., \& {Cooray}, A. 2010, \mnras, 406, 2421, \dodoi{10.1111/j.1365-2966.2010.16898.x}

\bibitem[{{Schaeffer} {et~al.}(2023){Schaeffer}, {Giri}, \& {Schneider}}]{Schaeffer23}
{Schaeffer}, T., {Giri}, S.~K., \& {Schneider}, A. 2023, arXiv e-prints, arXiv:2305.15466, \dodoi{10.48550/arXiv.2305.15466}

\bibitem[{{Schneider} {et~al.}(2023){Schneider}, {Schaeffer}, \& {Giri}}]{Schneider23}
{Schneider}, A., {Schaeffer}, T., \& {Giri}, S.~K. 2023, \prd, 108, 043030, \dodoi{10.1103/PhysRevD.108.043030}

\bibitem[{Schwartz(1965)}]{Schwartz1965OnBP}
Schwartz, L. 1965, Zeitschrift f{\"u}r Wahrscheinlichkeitstheorie und Verwandte Gebiete, 4, 10

\bibitem[{{Sellentin} {et~al.}(2014){Sellentin}, {Quartin}, \& {Amendola}}]{DALI}
{Sellentin}, E., {Quartin}, M., \& {Amendola}, L. 2014, \mnras, 441, 1831, \dodoi{10.1093/mnras/stu689}

\bibitem[{{Shaver} {et~al.}(1999){Shaver}, {Windhorst}, {Madau}, \& {de Bruyn}}]{Shaver99}
{Shaver}, P.~A., {Windhorst}, R.~A., {Madau}, P., \& {de Bruyn}, A.~G. 1999, \aap, 345, 380, \dodoi{10.48550/arXiv.astro-ph/9901320}

\bibitem[{{Shen} {et~al.}(2022){Shen}, {Anstey}, {de Lera Acedo}, \& {Fialkov}}]{Shen22}
{Shen}, E., {Anstey}, D., {de Lera Acedo}, E., \& {Fialkov}, A. 2022, \mnras, 515, 4565, \dodoi{10.1093/mnras/stac1900}

\bibitem[{{Sims} \& {Pober}(2020)}]{SimsPober20}
{Sims}, P.~H., \& {Pober}, J.~C. 2020, \mnras, 492, 22, \dodoi{10.1093/mnras/stz3388}

\bibitem[{{Singh} {et~al.}(2018){Singh}, {Subrahmanyan}, {Udaya Shankar}, {Sathyanarayana Rao}, {Fialkov}, {Cohen}, {Barkana}, {Girish}, {Raghunathan}, {Somashekar}, \& {Srivani}}]{SARAS2}
{Singh}, S., {Subrahmanyan}, R., {Udaya Shankar}, N., {et~al.} 2018, \apj, 858, 54, \dodoi{10.3847/1538-4357/aabae1}

\bibitem[{{Singh} {et~al.}(2022){Singh}, {Jishnu}, {Subrahmanyan}, {Udaya Shankar}, {Girish}, {Raghunathan}, {Somashekar}, {Srivani}, \& {Sathyanarayana Rao}}]{SARAS3}
{Singh}, S., {Jishnu}, N.~T., {Subrahmanyan}, R., {et~al.} 2022, Nature Astronomy, 6, 607, \dodoi{10.1038/s41550-022-01610-5}

\bibitem[{{Skilling}(2004)}]{Skilling04}
{Skilling}, J. 2004, in American Institute of Physics Conference Series, Vol. 735, Bayesian Inference and Maximum Entropy Methods in Science and Engineering: 24th International Workshop on Bayesian Inference and Maximum Entropy Methods in Science and Engineering, ed. R.~{Fischer}, R.~{Preuss}, \& U.~V. {Toussaint}, 395--405, \dodoi{10.1063/1.1835238}

\bibitem[{{Speagle}(2020)}]{Speagle20}
{Speagle}, J.~S. 2020, \mnras, 493, 3132, \dodoi{10.1093/mnras/staa278}

\bibitem[{{Tauscher} {et~al.}(2020){Tauscher}, {Rapetti}, \& {Burns}}]{Tauscher20}
{Tauscher}, K., {Rapetti}, D., \& {Burns}, J.~O. 2020, \apj, 897, 132, \dodoi{10.3847/1538-4357/ab9a3f}

\bibitem[{{Tauscher} {et~al.}(2021){Tauscher}, {Rapetti}, {Nhan}, {Handy}, {Bassett}, {Hibbard}, {Bordenave}, {Bradley}, \& {Burns}}]{PaperIV}
{Tauscher}, K., {Rapetti}, D., {Nhan}, B.~D., {et~al.} 2021, \apj, 915, 66, \dodoi{10.3847/1538-4357/ac00af}

\bibitem[{{The HERA Collaboration} {et~al.}(2022){The HERA Collaboration}, {Abdurashidova}, {Adams}, {Aguirre}, {Alexander}, {Ali}, {Baartman}, {Balfour}, {Barkana}, {Beardsley}, {Bernardi}, {Billings}, {Bowman}, {Bradley}, {Breitman}, {Bull}, {Burba}, {Carey}, {Carilli}, {Cheng}, {Choudhuri}, {DeBoer}, {de Lera Acedo}, {Dexter}, {Dillon}, {Ely}, {Ewall-Wice}, {Fagnoni}, {Fialkov}, {Fritz}, {Furlanetto}, {Gale-Sides}, {Garsden}, {Glendenning}, {Gorce}, {Gorthi}, {Greig}, {Grobbelaar}, {Halday}, {Hazelton}, {Heimersheim}, {Hewitt}, {Hickish}, {Jacobs}, {Julius}, {Kern}, {Kerrigan}, {Kittiwisit}, {Kohn}, {Kolopanis}, {Lanman}, {La Plante}, {Lewis}, {Liu}, {Loots}, {Ma}, {MacMahon}, {Malan}, {Malgas}, {Malgas}, {Maree}, {Marero}, {Martinot}, {McBride}, {Mesinger}, {Mirocha}, {Molewa}, {Morales}, {Mosiane}, {Mu{\~n}oz}, {Murray}, {Nagpal}, {Neben}, {Nikolic}, {Nunhokee}, {Nuwegeld}, {Parsons}, {Pascua}, {Patra}, {Pieterse}, {Qin}, {Razavi-Ghods}, {Robnett}, {Rosie}, {Santos}, {Sims}, {Singh}, {Smith}, {Swarts},
  {Tan}, {Thyagarajan}, {Wilensky}, {Williams}, {van Wyngaarden}, \& {Zheng}}]{HERA}
{The HERA Collaboration}, {Abdurashidova}, Z., {Adams}, T., {et~al.} 2022, arXiv e-prints, arXiv:2210.04912, \dodoi{10.48550/arXiv.2210.04912}

\bibitem[{{Thomas} {et~al.}(2009){Thomas}, {Zaroubi}, {Ciardi}, {Pawlik}, {Labropoulos}, {Jeli{\'c}}, {Bernardi}, {Brentjens}, {de Bruyn}, {Harker}, {Koopmans}, {Mellema}, {Pandey}, {Schaye}, \& {Yatawatta}}]{Thomas09}
{Thomas}, R.~M., {Zaroubi}, S., {Ciardi}, B., {et~al.} 2009, \mnras, 393, 32, \dodoi{10.1111/j.1365-2966.2008.14206.x}

\bibitem[{{Trott} {et~al.}(2020){Trott}, {Jordan}, {Midgley}, {Barry}, {Greig}, {Pindor}, {Cook}, {Sleap}, {Tingay}, {Ung}, {Hancock}, {Williams}, {Bowman}, {Byrne}, {Chokshi}, {Hazelton}, {Hasegawa}, {Jacobs}, {Joseph}, {Li}, {Line}, {Lynch}, {McKinley}, {Mitchell}, {Morales}, {Ouchi}, {Pober}, {Rahimi}, {Takahashi}, {Wayth}, {Webster}, {Wilensky}, {Wyithe}, {Yoshiura}, {Zhang}, \& {Zheng}}]{MWA}
{Trott}, C.~M., {Jordan}, C.~H., {Midgley}, S., {et~al.} 2020, \mnras, 493, 4711, \dodoi{10.1093/mnras/staa414}

\bibitem[{{Trotta}(2008)}]{Trotta08}
{Trotta}, R. 2008, Contemporary Physics, 49, 71, \dodoi{10.1080/00107510802066753}

\bibitem[{Van~Rossum \& Drake~Jr(1995)}]{python}
Van~Rossum, G., \& Drake~Jr, F.~L. 1995, Python reference manual (Centrum voor Wiskunde en Informatica Amsterdam)

\bibitem[{{Virtanen} {et~al.}(2020){Virtanen}, {Gommers}, {Oliphant}, {Haberland}, {Reddy}, {Cournapeau}, {Burovski}, {Peterson}, {Weckesser}, {Bright}, {van der Walt}, {Brett}, {Wilson}, {Millman}, {Mayorov}, {Nelson}, {Jones}, {Kern}, {Larson}, {Carey}, {Polat}, {Feng}, {Moore}, {VanderPlas}, {Laxalde}, {Perktold}, {Cimrman}, {Henriksen}, {Quintero}, {Harris}, {Archibald}, {Ribeiro}, {Pedregosa}, {van Mulbregt}, \& {SciPy 1. 0 Contributors}}]{scipy}
{Virtanen}, P., {Gommers}, R., {Oliphant}, T.~E., {et~al.} 2020, Nature Methods, 17, 261, \dodoi{10.1038/s41592-019-0686-2}

\bibitem[{{Zhang} {et~al.}(2022){Zhang}, {Shan}, {Gu}, {Zheng}, {Xu}, {Yue}, {Liu}, {Zhu}, \& {Guo}}]{Zhang22}
{Zhang}, Z., {Shan}, H., {Gu}, J., {et~al.} 2022, \mnras, 516, 1573, \dodoi{10.1093/mnras/stac2208}

\end{thebibliography}
\bibliographystyle{aasjournal}
\end{document}